\documentclass[final,5p,times,twocolumn]{elsarticle}
\usepackage{amssymb}
\usepackage{amsmath}
\usepackage{subcaption}
\usepackage{xcolor}
\usepackage{float}
\usepackage{booktabs} 
\usepackage[hidelinks]{hyperref}
\biboptions{numbers,sort&compress} % 启用编号、排序和压缩连续编号
\setcitestyle{square} % 设置引用标号为方括号
\usepackage[none]{hyphenat}
\usepackage[switch]{lineno}

\journal{Nuclear Instruments and Methods in Physics Research Section A}

\begin{document}
% \linenumbers
\begin{frontmatter}



\title{A Domain Adaptive Position Reconstruction Method for Time Projection Chamber based on Deep Neural Network} %% Article title


\author[label1,label2,label3]{Xiaoran Guo}

\cortext[cor1]{Corresponding author.}
\author[label4]{Fei Gao}
\author[label4]{Kaihang Li}
\author[label1,label2,label3]{Qing Lin\corref{cor1}}
\ead{qinglin@ustc.edu.cn}
\author[label5]{Jiajun Liu}
\author[label1,label2,label3]{Lijun Tong}
\author[label5]{Xiang Xiao}
\author[label4]{Lingfeng Xie}
\author[label4]{Yifei Zhao}

\affiliation[label1]{organization={State Key Laboratory of Particle Detection and Electronics, University of Science and Technology of China},
            city={Hefei},
         postcode={230026},
            country={China}}

\affiliation[label2]{organization={Department of Modern Physics, University of Science and Technology of China},
           city={Hefei},
          postcode={230026},
             country={China}}
\affiliation[label3]{organization={Deep Space Exploration Laboratory},
           city={Hefei},
          postcode={230026},
             country={China}}
        
\affiliation[label4]{organization={Department of Physics \& Center for High Energy Physics, Tsinghua University}, city={Beijing},postcode={100084}, country={China}}
\affiliation[label5]{organization={School of Physics, Sun Yat-Sen University}, city={Guangzhou},postcode={510275}, country={China}}
%% Abstract
\begin{abstract}

Transverse position reconstruction in a Time Projection Chamber (TPC) is crucial for accurate particle tracking and classification, and is typically accomplished using machine learning techniques. However, these methods often exhibit biases and limited resolution due to incompatibility between real experimental data and simulated training samples. To mitigate this issue, we present a domain-adaptive reconstruction approach based on a cycle-consistent generative adversarial network.  In the prototype detector, the application of this method led to a 60.6\% increase in the reconstructed radial boundary. Scaling this method to a simulated 50-kg TPC, by evaluating the resolution of simulated events, an additional improvement of at least 27\% is achieved.
\end{abstract}


% \begin{highlights}
% \end{highlights}

%% Keywords

\end{frontmatter}



\section{Introduction}

Time Projection Chambers (TPCs) are widely used in particle physics experiments due to their capability for high-resolution three-dimensional event reconstruction. In dual-phase TPCs, the reconstruction of the XY-position primarily relies on analyzing signals collected by the top photomultiplier tube (PMT) array. Accurate reconstruction of event positions is essential for enhancing the detector’s overall sensitivity and improving background rejection, thus often used for dark matter direct searches  ~\citep{bo2025dark,xenon2024xenonnt,akerib2020lux,agnes2018low,agnes2023sensitivity,xu2019electron}.

In recent years, machine learning—particularly deep neural networks—has been increasingly adopted for signal tasks in particle detectors ~\citep{Akerib2018,Delaquis2018,Aprile2025,gong2019machine,adams2019deep,lobasenko2023reconstruction,koh2023deep}. However, a persistent challenge arises incompatibility between the simulated data used for training and the experimental data encountered during inference. This distributional mismatch, often referred to as domain shift, can degrade model performance when deployed in real-world settings, especially in detector regions that are poorly represented in the training data.

To mitigate this issue, Domain Adaptation (DA) techniques have been proposed to reduce the distribution gap between the source and target domains. These methods aim to transfer knowledge from labeled synthetic data to unlabeled or sparsely labeled real data by aligning their feature representations. A theoretical foundation for this approach was introduced by Ben David et al. ~\citep{BenDavid2010}, and it has since become central to research in this area.

Generative Adversarial Networks (GANs) provide a powerful framework for generating synthetic data that mimics the distribution of real data ~\citep{Goodfellow2014}. Building upon this, Cycle-Consistent Generative Adversarial Networks (CycleGANs) were proposed ~\citep{Zhu2017} to enable unsupervised image-to-image translation between two domains without requiring paired training data. Similar unsupervised techniques have demonstrated strong cross-domain generation capabilities in fields such as computer vision ~\citep{Yi2017,csurka2017domain,xu2019self,liu2022deep,oza2023unsupervised}, medical image synthesis ~\citep{Yang2018,guan2021domain}, and artistic style transfer ~\citep{Goodfellow2014,pizzati2020domain,luo2020adversarial}. Such methods can be particularly advantageous in high-energy physics, where obtaining labeled experimental data is extremely challenging. CycleGANs have become a powerful framework for unpaired image-to-image translation tasks. Unlike traditional supervised methods that rely on precisely aligned input-output pairs, CycleGANs enable bidirectional domain translation without requiring such one-to-one correspondence by integrating adversarial training with a cycle consistency constraint. This feature makes CycleGANs particularly suitable for applications in physical experiments, where perfectly paired simulation and experimental data are often unavailable or difficult to obtain.

In this work, we propose a DA framework based on CycleGANs to translate simulated data into a distribution that more closely resembles experimental data, thereby reducing bias in position reconstruction. We integrate this domain-adapted data into a Deep Residual Network ~\citep{he2016deep} for accurate XY-position reconstruction in dual-phase TPCs. By utilizing the generative capability of CycleGANs and the regression power of DeepResNet, our method offers improved robustness and accuracy under real detector conditions. 

The structure of this paper is as follows. Section 2 introduces the reconstruction methodology, including the CycleGAN-based DA used for unpaired simulation to experiment translation, as well as the ResNet model designed for position regression. Section 3 describes the signal simulation procedure and the construction of the training and evaluation datasets. Section 4 presents the results of the proposed reconstruction method, including its expected performance on RELICS experiment  and experimental validation using data from the RELICS prototype.

\label{sec1}

\section{Position Reconstruction Algorithm}
\begin{figure*}[!htb]
    \centering
    \includegraphics[width=1\textwidth]{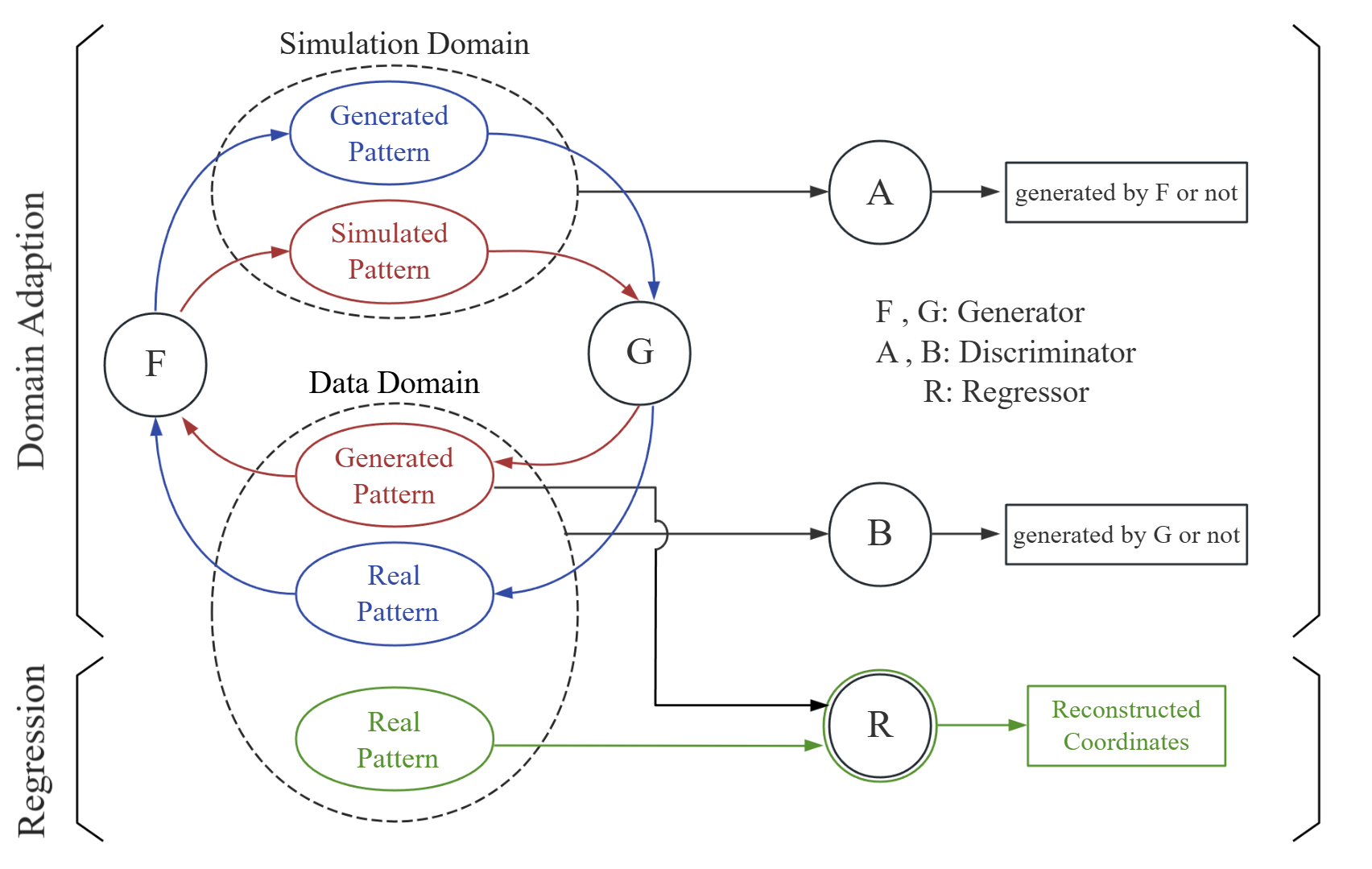}
    \caption{Schematic illustration of the training framework. 
    The upper part (Domain Adaption stage) establishes mappings between the simulation domain and the data domain through two Generators F and G. Their outputs are evaluated by discriminators A and B, which attempt to distinguish between the generator-produced samples and the real samples belonging to the corresponding domain.The blue and red arrows represent two training cycles. The black arrows represent other training processes.
    The lower part (Regression stage) uses paired coordinates together with the transformed data as input to a regressor for position reconstruction.  The green arrows represent the inference process.
    }
    \label{fig:workflow}
\end{figure*}

The overall workflow is illustrated in Fig.~\ref{fig:workflow}. We begin by training a DA model using both simulated and data domain. In our work, the simulation domain contains Monte Carlo-generated signals, which will be described in detail in \autoref{signal}, whereas data domain, containing signals derived from experimental measurements or detector prototypes. This model learns a mapping that transforms samples from the simulation domain to resemble those in the data domain. Once trained, it is applied to convert large volumes of simulation data into a domain-adapted dataset. This adapted dataset is then used to train a position reconstruction model, which is subsequently evaluated on real experimental data. This ensures that the reconstruction model is exposed to data distributions similar to those observed in practice, thereby enhancing its generalization performance and reducing the risk of systematic errors induced by domain shift.

\subsection{CycleGAN for Domain Adaptation}

As depicted in \autoref{fig:workflow}, a CycleGAN  consists of two generators. The generator G translates patterns from the simulation domain to the data domain, while F performs the reverse transformation, along with two discriminators, A and B . The generators are tasked with translating samples between the two domains, while the discriminators aim to distinguish real samples from generated ones within their respective domains. 

To prevent the generators from producing outputs that are visually plausible yet physically inconsistent, CycleGAN employs a cycle consistency loss that enforces reversibility of the mappings. Concretely, an input \( x \) from simulation domain translated to data domain by \( G \), and then mapped back to simulation domain by \( F \), should closely reconstruct the original input, i.e., \( F(G(x)) \approx x \). The same constraint is applied symmetrically for inputs from the other domain. This loss acts as a powerful regularizer, preserving the underlying structural and semantic features during domain translation. By jointly optimizing the adversarial losses and cycle consistency loss, CycleGAN learns meaningful and physically consistent cross-domain mappings. After training, the generator \( G \) can effectively translate simulated inputs into realistic data-like distributions, thereby reducing the domain gap before applying subsequent regression networks for accurate position reconstruction.

\begin{figure}[!ht]
    \centering
    \includegraphics[width=1\linewidth]{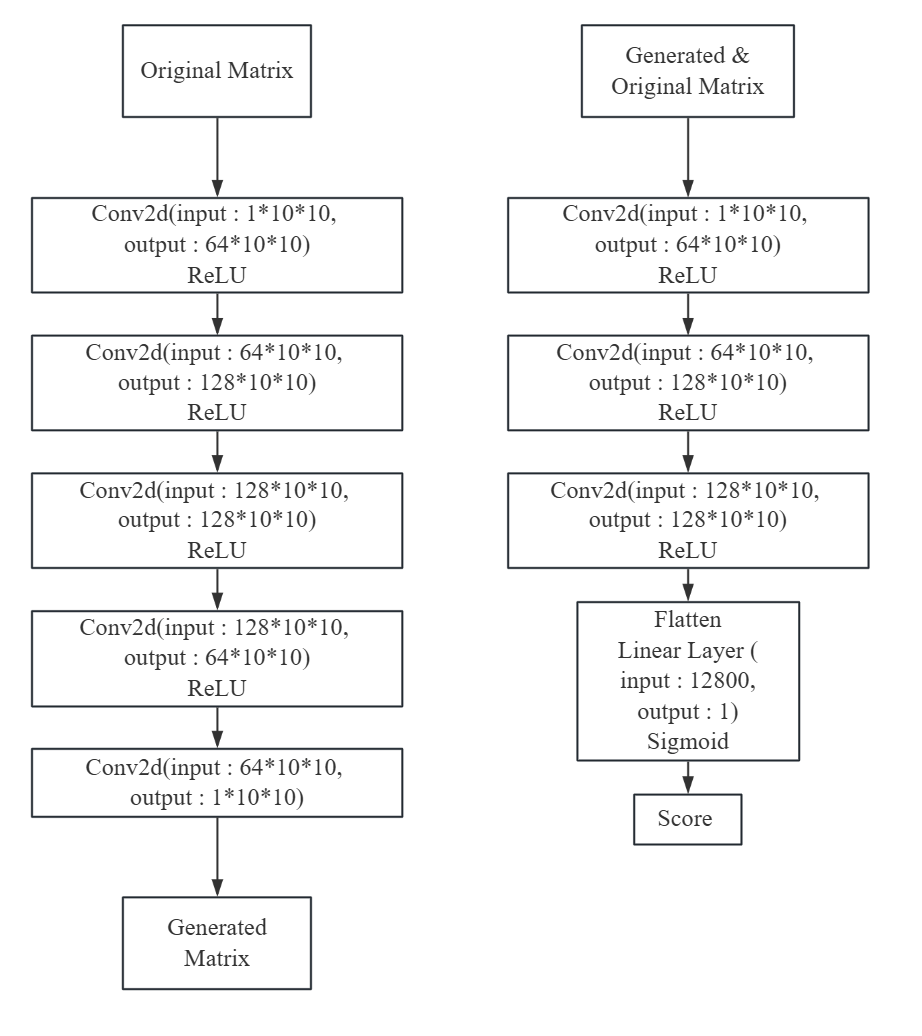}
    \caption{The left panel illustrates the architecture of the generators \( G \) and \( F \), which are responsible for translating data between two domains in the CycleGAN framework. The right panel shows the structure of the discriminators \( D_X \) and \( D_Y \), which are used to distinguish between real and generated samples. A discriminator outputs a score between 0 and 1; outputs greater than 0.5 are classified as real, while outputs less than 0.5 are considered fake.}
    \label{fig:architectCGAN}
\end{figure}

\begin{figure}[!ht]
    \centering
    \includegraphics[width=1\linewidth]{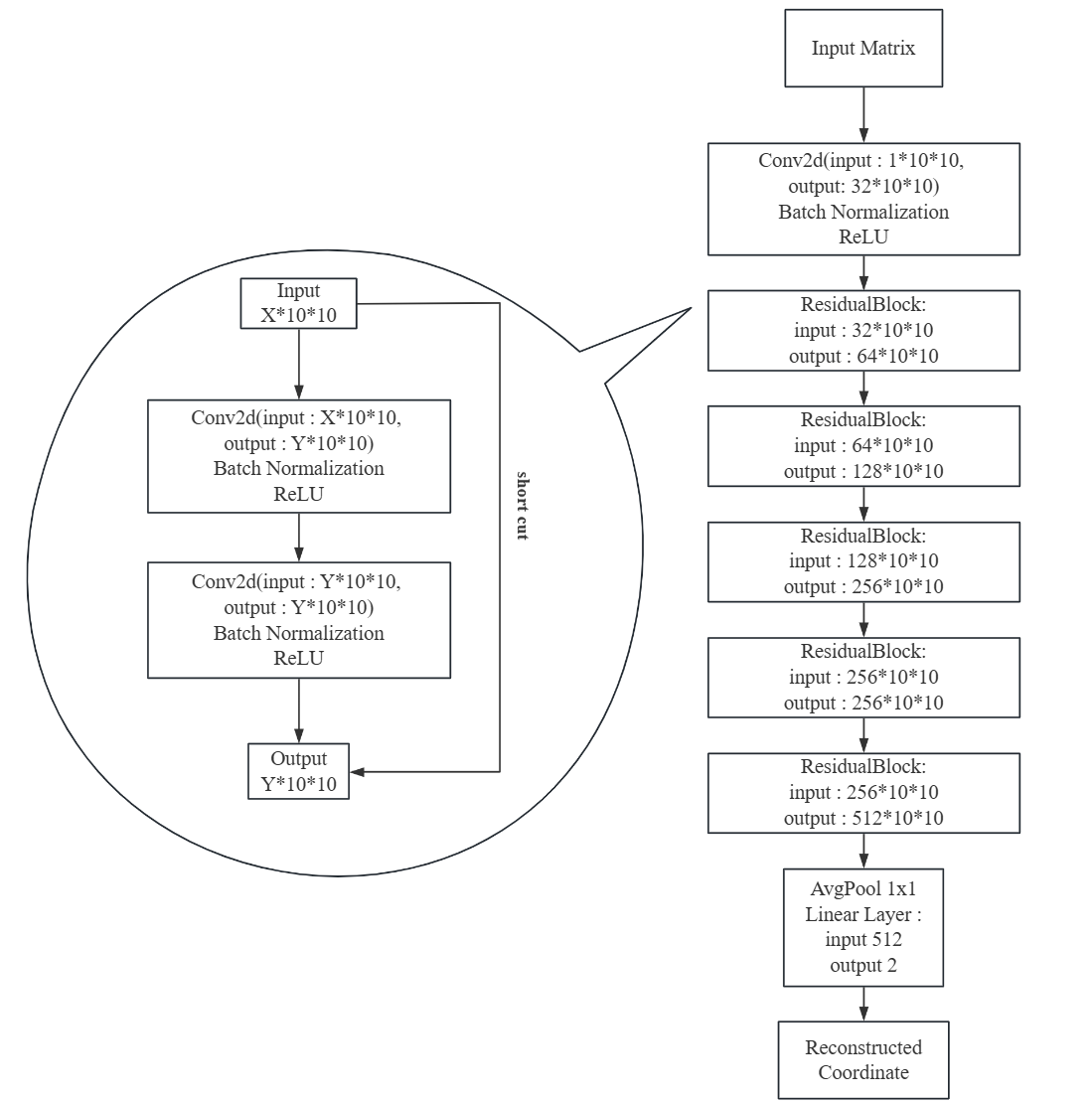}
    \caption{The right panel shows the architecture of the Deep Residual Network, while the left illustrates the detailed structure of a single residual block. The input and output channels \(X\) and \(Y\) are adjusted according to the channel labels indicated in the network diagram on the right.A shortcut connection is added in each residual block to mitigate gradient vanishing and improve overall network performance.}
    \label{fig:DRNarchitect}
\end{figure}

During training, the optimization is performed in an alternating manner. First, the discriminators \(A\) and \(B\) are updated to minimize their losses \(\mathcal{L}_A\) and \(\mathcal{L}_B\), thereby improving their ability to distinguish real samples from generated ones. Next, with the discriminators fixed, the generators \(G\) and \(F\) are updated to minimize the total loss \(\mathcal{L}_{G,F}\). This iterative procedure enables the generators to gradually learn mappings that not only fool the discriminators but also preserve the structural consistency across domains through the cycle-consistency constraint.

The total generator loss combines adversarial and cycle consistency terms:

\begin{equation}
\mathcal{L}_{G,F} = \mathcal{L}_{\text{adv}}(G) + \mathcal{L}_{\text{adv}}(F) + \lambda \cdot \mathcal{L}_{\text{cyc}},
\end{equation}
where \(\lambda = 90\) balances the adversarial loss and the cycle consistency loss.

The training of CycleGAN relies on both adversarial losses for the generators and discriminators, as well as the cycle consistency loss. The adversarial losses for the generators are defined as:

\begin{equation}
\mathcal{L}_{\text{adv}}(G) = \mathbb{E}_{x \sim p_{\text{sim}}(x)} \big[ \ell_{\text{BCE}}(B(G(x)), 1) \big]
\end{equation}

\begin{equation}
\mathcal{L}_{\text{adv}}(F) = \mathbb{E}_{y \sim p_{\text{data}}(y)} \big[ \ell_{\text{BCE}}(A(F(y)), 1) \big]
\end{equation}
where $\ell_{\text{BCE}}(p, t) = -[\, t \log(p) + (1 - t) \log(1 - p) \,]$ denotes the binary cross-entropy loss, 
 $p$ represents the discriminator's predicted probability that a given input is real, and 
$t \in \{0, 1\}$ is the corresponding ground-truth label indicating whether the input is real ($t=1$) or generated ($t=0$). 
The discriminators $A$ and $B$ are associated with the data and simulation domains, respectively. 
These adversarial objectives encourage the generators $G$ and $F$ to produce outputs that are indistinguishable from real samples in their corresponding target domains.
\(x \sim p_{\text{sim}}(x)\) represents a sample from the simulation domain,  
\(y \sim p_{\text{data}}(y)\) represents a sample from the data domain,  
\(G(x)\) and \(F(y)\) are the generated samples from generators \(G\) and \(F\), respectively,  
\(B(\cdot)\) and \(A(\cdot)\) are the scores given by discriminators \(B\) and \(A\), representing the confidence that a sample belongs to the real domain,  
and 1 indicates that the generator aims to fool the discriminator into classifying the fake sample as real. The expectations \(\mathbb{E}[\cdot]\) are approximated by mini-batch averages during training, with a batch size of 128.

The cycle consistency loss is given by

\begin{equation}
\begin{split}
\mathcal{L}_{\text{cyc}} =& \mathbb{E}_{x \sim p_{\text{sim}}(x)} \big[ \| F(G(x)) - x \|_1 \big] + \\
&  \mathbb{E}_{y \sim p_{\text{data}}(y)} \big[ \| G(F(y)) - y \|_1 \big]
\end{split}
\end{equation}
where \(\|v\|_1 = \sum_i |v_i|\) denotes the \(L_1\) norm, which measures the element-wise difference between the original sample and its reconstruction. The same procedure applies for data domain samples.

The discriminator losses are defined as

\begin{equation}
\begin{split}
\mathcal{L}_{A} =& \mathbb{E}_{x \sim p_{\text{sim}}(x)} \big[ \ell_{\text{BCE}}(A(x), 1) \big] + \\
&\mathbb{E}_{y \sim p_{\text{data}}(y)} \big[ \ell_{\text{BCE}}(A(F(y)), 0) \big]  
\end{split}
\end{equation}

\begin{equation}
\begin{split}
\mathcal{L}_{B} =& \mathbb{E}_{y \sim p_{\text{data}}(y)} \big[ \ell_{\text{BCE}}(B(y), 1) \big] + \\
&\mathbb{E}_{x \sim p_{\text{sim}}(x)} \big[ \ell_{\text{BCE}}(B(G(x)), 0) \big]  
\end{split}
\end{equation}
where 0 indicates that the discriminator should classify generated samples as fake, and 1 indicates classification of real samples as real.

\subsection{Deep Residual Network for Position Regression}

Following the DA step, in which data from the simulation domain is transformed to better align with the characteristics of the data domain, we employ a Residual Network (ResNet) to regress the spatial coordinates of detected events. The ResNet architecture, illustrated in \autoref{fig:DRNarchitect}, is composed of multiple residual blocks that enable the training of deeper networks by facilitating the flow of gradients through shortcut connections.

The input to the ResNet is the adapted simulated data produced by the CycleGAN, which now shares the distributional properties of the data domain and preserves subtle features crucial for precise position reconstruction. By exploiting the hierarchical feature extraction capabilities of deep convolutional layers combined with residual learning, the ResNet can effectively model the complex nonlinear mapping between the input signal patterns and the corresponding spatial positions.

For training, we adopt the Smooth L1 loss function (also known as the Huber loss), defined as:

\begin{equation}
\mathcal{L}_{smooth}(x, y) =
\begin{cases}
0.5 \, \|x - y\|_2^2, & \text{if } \|x - y\|_1 < 1 \\
\|x - y\|_1 - 0.5, & \text{otherwise}
\end{cases}
\label{smoothl1}
\end{equation}
Here, \(\|v\|_1 = \sum_i |v_i|\) denotes the L1 norm, and \(\|v\|_2^2 = \sum_i v_i^2\) denotes the squared L2 norm. The Smooth L1 loss combines the advantages of the L2 loss and the L1 loss: for small errors (\(\|x - y\|_1 < 1\)), it behaves like the L2 loss, providing smooth gradients that encourage accurate fits; for larger errors, it transitions to the L1 loss, which is more robust against outliers and helps stabilize training by limiting excessively large gradient updates. Such properties make the Smooth L1 loss especially suitable for regression problems in experimental physics, where measurement noise and label uncertainty are common.

\section{Signal simulation}
\label{signal}
% \subsection{Charge Amplification}
% In the liquid xenon region, electron-ion pairs generated during ionization undergo recombination and emit scintillation light. A fraction of electrons escape recombination and drift under the electric field towards the liquid surface. When these electrons reach the liquid-gas interface, they are extracted into the gaseous xenon region by a high electric field. The escaped electrons then produce secondary scintillation through proportional amplification in the gas phase, which is termed the proportional scintillation signal---S2\citep{aprile2011design}.

When a particle deposits energy in the active medium of a TPC, both excitation and ionization of the target atoms or molecules occur. The recombination of electron–ion pairs gives rise to an immediate flash of scintillation light. A fraction of the liberated electrons escape recombination and drift under the applied electric field. Upon reaching the amplification region, these electrons are extracted and accelerated, inducing secondary scintillation or charge amplification, depending on the detector configuration. This delayed signal is spatially localized near the amplification region.

The relative timing between the prompt and delayed signals provides a measure of the electron drift time, which, together with the known drift velocity, determines the depth (Z coordinate) of the original interaction vertex along the drift direction. At the same time, the distribution of the delayed signal across the photosensor array reflects the lateral spread of the drifting electrons. By analyzing this pattern, the transverse coordinates (X and Y) of the interaction can be reconstructed. In this way, a TPC achieves full three-dimensional localization of each event.

Through dedicated signal simulations, one can predict how an energy deposition at a given transverse position (X,Y) translates into the response of the photosensor array. The optical model provides the expected detection probabilities for each sensor, and the resulting response pattern serves as a characteristic "fingerprint" of that interaction position. In the following discussion, liquid xenon TPCs are used as a representative example to illustrate this process.

 \subsection{Simulation procedures}

\begin{figure}[!htb]
    \centering
    % --------------------
    % 上图：XY平面
    % --------------------
    \begin{subfigure}[n]{\linewidth}
        \centering
        \includegraphics[width=\linewidth]{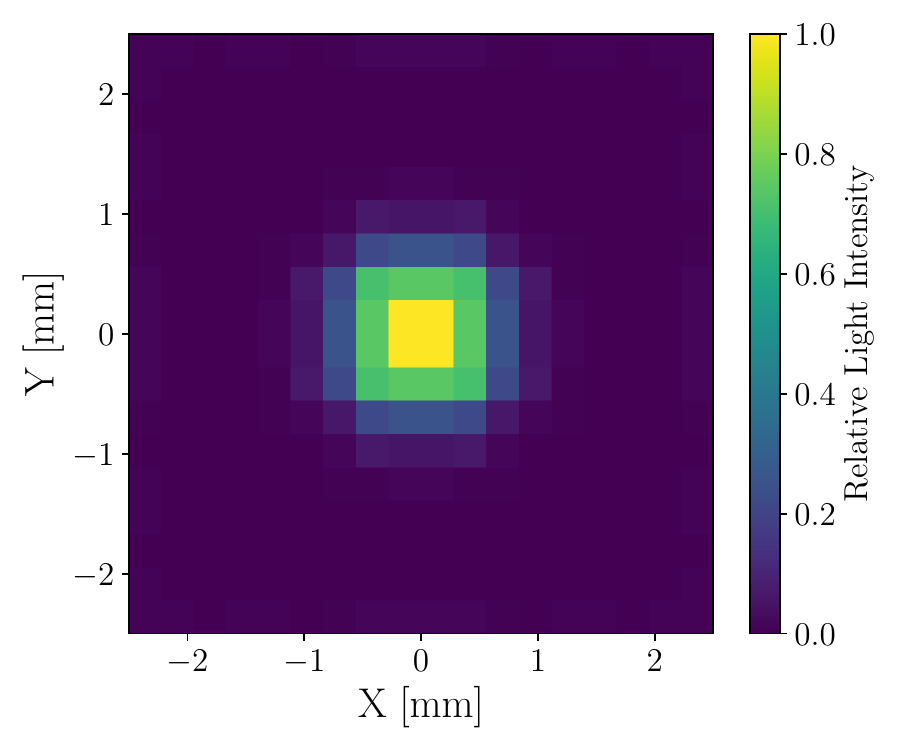}
        \caption{}
        \label{fig:xy_density}
    \end{subfigure}
    \vspace{0.1cm}
    % --------------------
    % 下图：ZX平面
    % --------------------
    \begin{subfigure}[n]{\linewidth}
        \centering
        \includegraphics[width=\linewidth]{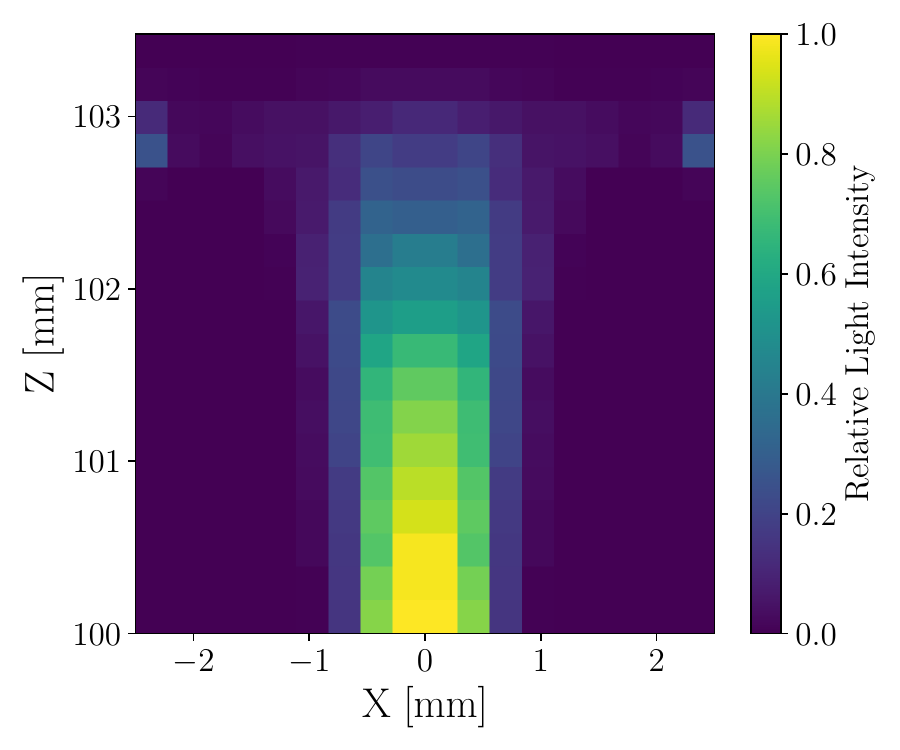}
        \caption{}
        \label{fig:zx_density}
    \end{subfigure}
    
    % --------------------
    % 总图例
    % --------------------
    \caption{(a) XY plane density; (b) ZX plane density. Relative scintillation intensity distribution in a single 5~mm $\times$ 5~mm electrode grid (maximum value = 1). 
    The color represents the normalized scintillation intensity in each plane. 
    This distribution is calculated for a single cell using Garfield++ based on the electric field simulated by COMSOL. 
    The strongest scintillation occurs at the center of the cell where the electric field is maximal. 
    Near the grid edges, scintillation can also appear due to the influence of the anode wires.}
    \label{fig:light_distribution}
\end{figure}

\begin{figure}[htbp]
    \centering
    \includegraphics[width=1\linewidth]{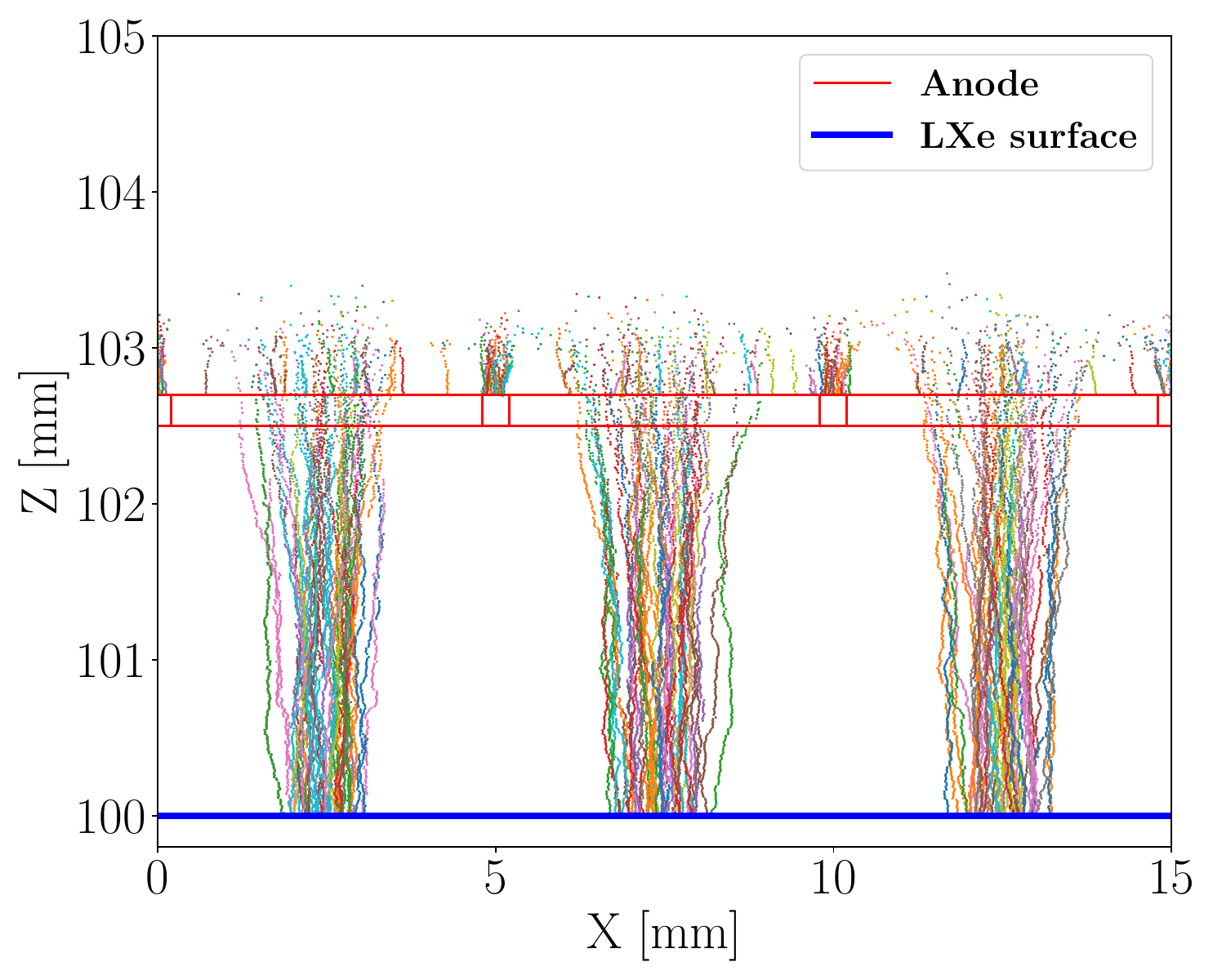}
    \caption{Scintillation simulated in gaseous Xenon. Each point of a certain color represents the location where an individual electron emits a photon in gaseous xenon, approximately tracing its path under the influence of the electric field. The blue horizontal line indicates the liquid surface level. The red region illustrates the shape of the anode wires, and each red rectangle spaced by 5 mm represents the position of an anode wire perpendicular to the X-axis, extending outward along the vertical coordinate system. The initial positions of the electrons in liquid xenon are uniformly distributed. }
    \label{fig:Garfield++}
\end{figure}

% The charge amplification process in the gaseous xenon region is simulated using Garfield++ ~\citep{schindler2018garfield++}, while the electric field configuration in the drift region is obtained through finite element simulations using COMSOL Multiphysics ~\citep{comsol}. Since the electrode wires form a regular mesh structure, we apply periodic boundary conditions in the field simulation to represent the repeating unit of the grid. This approximation allows us to model a single unit cell instead of the entire array, significantly reducing computational cost while preserving the electric field characteristics near the mesh. This simulation models the drift process of escaped electrons through both liquid and gaseous xenon phases, while recording the positions of electroluminescence events in the gas phase, as shown in \autoref{fig:Garfield++}. 

The charge amplification process in the gaseous xenon region is simulated using Garfield++ ~\citep{schindler2018garfield++}, while the electric field configuration in the drift region is obtained through finite element simulations using COMSOL Multiphysics ~\citep{comsol}.The resulting scintillation intensity distributions in a single cell are shown in \autoref{fig:xy_density} for the XY plane and \autoref{fig:zx_density} for the ZX plane. These distributions are based on the electric field simulated by COMSOL and calculated using Garfield++. The strongest scintillation occurs at the center of the cell where the electric field is maximal, while near the grid edges, scintillation can also appear due to the influence of the anode wires. 
In general, we first model the drift of all electrons generated by a single ionization event at a fixed horizontal location within one electrode grid in liquid xenon. A 100\% electron extraction efficiency is assumed at the liquid-gas interface. The spatial distribution of scintillation photons produced by these electrons in the gaseous xenon is recorded, later are used in optical simualtion. Owing to the periodic boundary conditions imposed on the electric field, the simulation results obtained from a single grid unit can be extrapolated to the full electrode plane using translational symmetry. 

The optical simulation is done by the simulation framework for the RELICS experiment \citep{Cai2024}, RelicsSim, which is based on BambooMC \citep{chen2021bamboomc} toolkit using GEANT4 \citep{agostinelli2003geant4}. The output data from Garfield++ simulations serves as light source input for this process\citep{relicsprototype}.  This approximation allows us to model a single unit cell instead of the entire array, significantly reducing computational cost while preserving the electric field characteristics near the mesh. This simulation models the drift process of escaped electrons through both liquid and gaseous xenon phases, while recording the positions of electroluminescence events in the gas phase, as shown in \autoref{fig:Garfield++}. A particle gun module implemented in RelicsSim then reads the simulated photon data and randomly samples from it to simulate photon propagation within the full detector geometry. The overall scintillation intensity can be tuned by adjusting the number of extracted electrons associated with each event.

\subsection{Detector readout geometries}
The RELICS experiment \citep{Cai2024} is planned to detect Coherent Elastic Neutrino-Nucleus Scattering (CE$\nu$NS) using Liquid Xenon Time Projection Chamber (LXeTPC). To validate the experimental design and detector performance, the Tsinghua University experimental team is currently operating a prototype of the detector\citep{relicsprototype}. This prototype will provide essential insights and help optimize the final design of the RELICS experiment. 

\begin{figure}[htb]
    \centering
    \includegraphics[width=\linewidth]{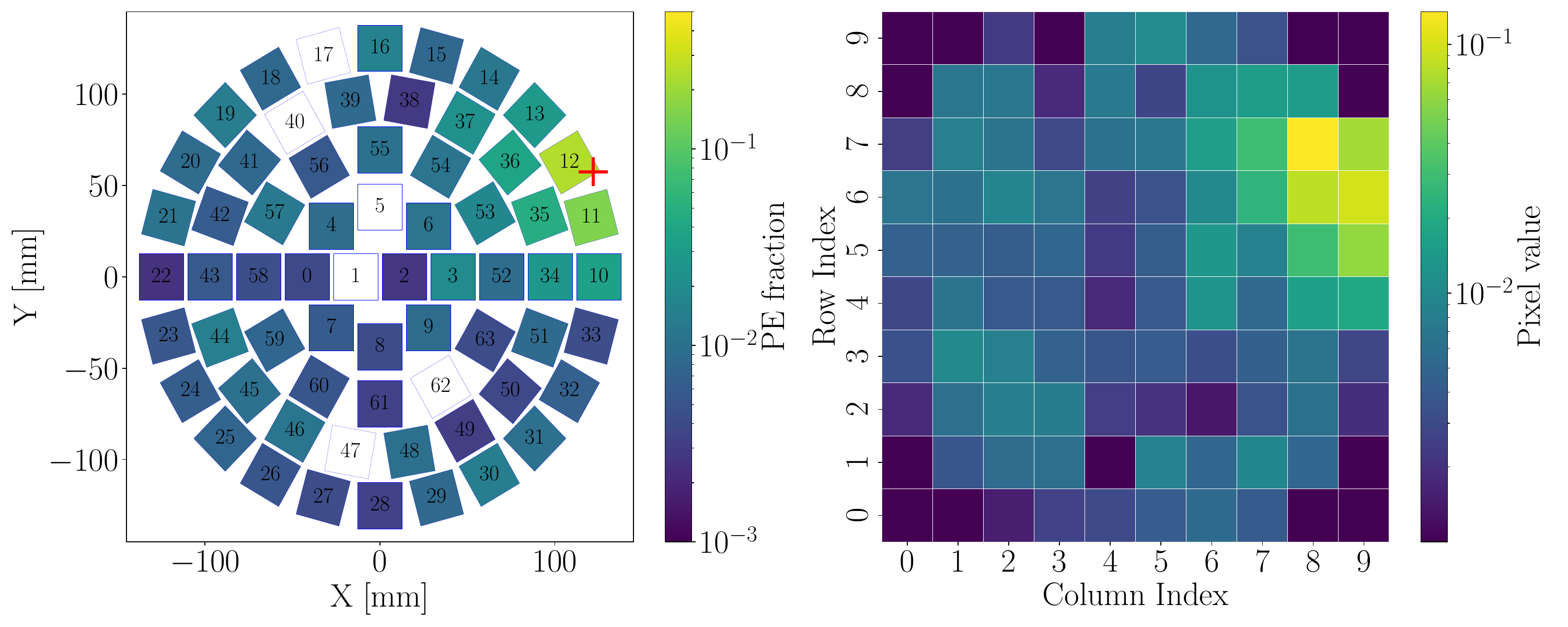}
    \caption{Transform a simulated PMT pattern into The \(10 \times 10\) matrix used to train the network. The red cross is the real position of the event, the XY coordinate of is the label for training. Due to the fact that the prototype detector’s PMT array consists of only 7 PMTs, a smaller \(4 \times 4\) matrix is used to represent it, based on the actual physical area of the array, using a similar mapping method. The training set contains 80,000 events, and tested on a 20,000 testing set. Finally validated on a 100,000 validation set. Each pattern represents a event which the top PMT array detect an average photon number of 800. }
    \label{fig:data_trans}
\end{figure}

%% Use \subsection commands to start a subsection.
RELICS detector is designed as a cylindrical TPC of 24\,cm in height and 28\,cm in diameter sits in a double-layer stainless steel (S.S) vessel, which provides a stable cryogenic environment. The target volume containing an active LXe of approximately 44\,kg is defined by 12 interlocking and light-tight PTFE (polytetrafluoroethylene, Teflon) panels. The active LXe is then viewed by two arrays of 64 Hamamatsu R8520-406 PMTs on top and bottom. \autoref{fig:data_trans} shows a typical PMT pattern generated by our simulation framework.
% \begin{figure}[htbp]
%     \centering
%     \includegraphics[width=1\linewidth]{fig1_relics_tpc.pdf}
%     \caption{The RELICS detector with the inner S.S. vessel, diving bell, veto PMTs, and the center TPC consisting of the top and bottom PMT arrays, anode, gate and cathode electrodes, copper field shaping-rings and PTFE reflector.}
%     \label{fig:relics_tpc}
% \end{figure}

\begin{figure}[htb]
    \centering
    \includegraphics[width=0.8\linewidth]{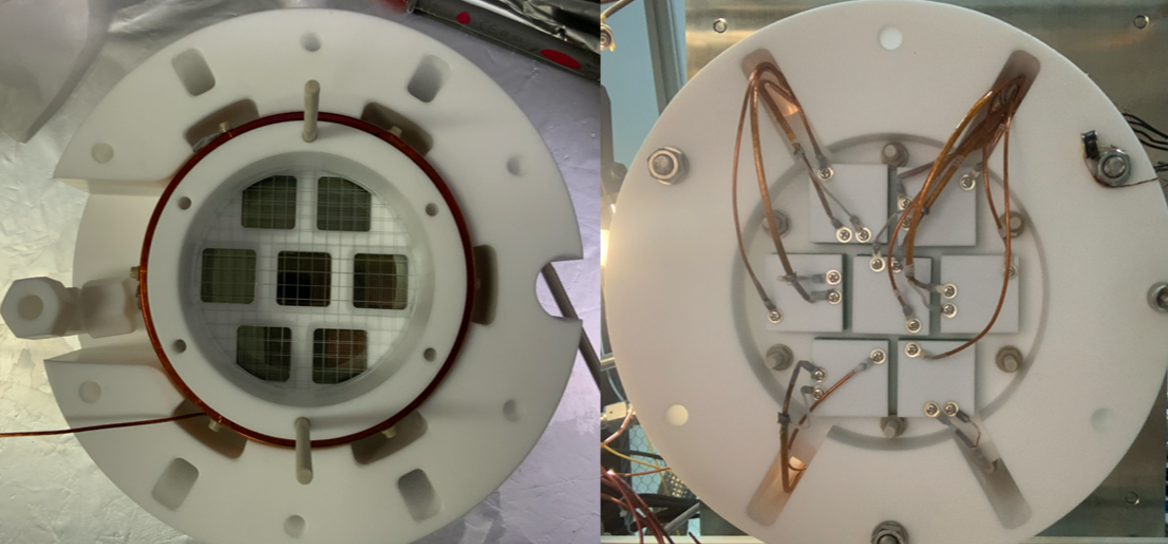}
    \caption{Photographs of the PMT array used in RELICS prototype, taken while assembling the detector in Tsinghua University.}
    \label{fig:prototype_assemble}
\end{figure}

The prototype for RELICS is a more compact TPC. It is a cylindrical of 3.6\,cm in height and 8\,cm in diameter.  The two arrays of 7 PMTs for each on top and bottom, which uses the same type of PMT for RELICS design. The arrangement of the PMT array is shown in \autoref{fig:prototype_assemble}.

% \begin{figure}[htbp]
%     \centering
%     \includegraphics[width=1\linewidth]{prototypedesign.png}
%     \caption{The prototype detector is basically a shrink version of RELICS detector, but without veto PMTs.Those photos shows the PMT array used in the RELICS prototype.}
%     \label{fig:prototype_tpc}
% \end{figure}

\begin{figure}[htb]
    \centering
    \includegraphics[width=0.8\linewidth]{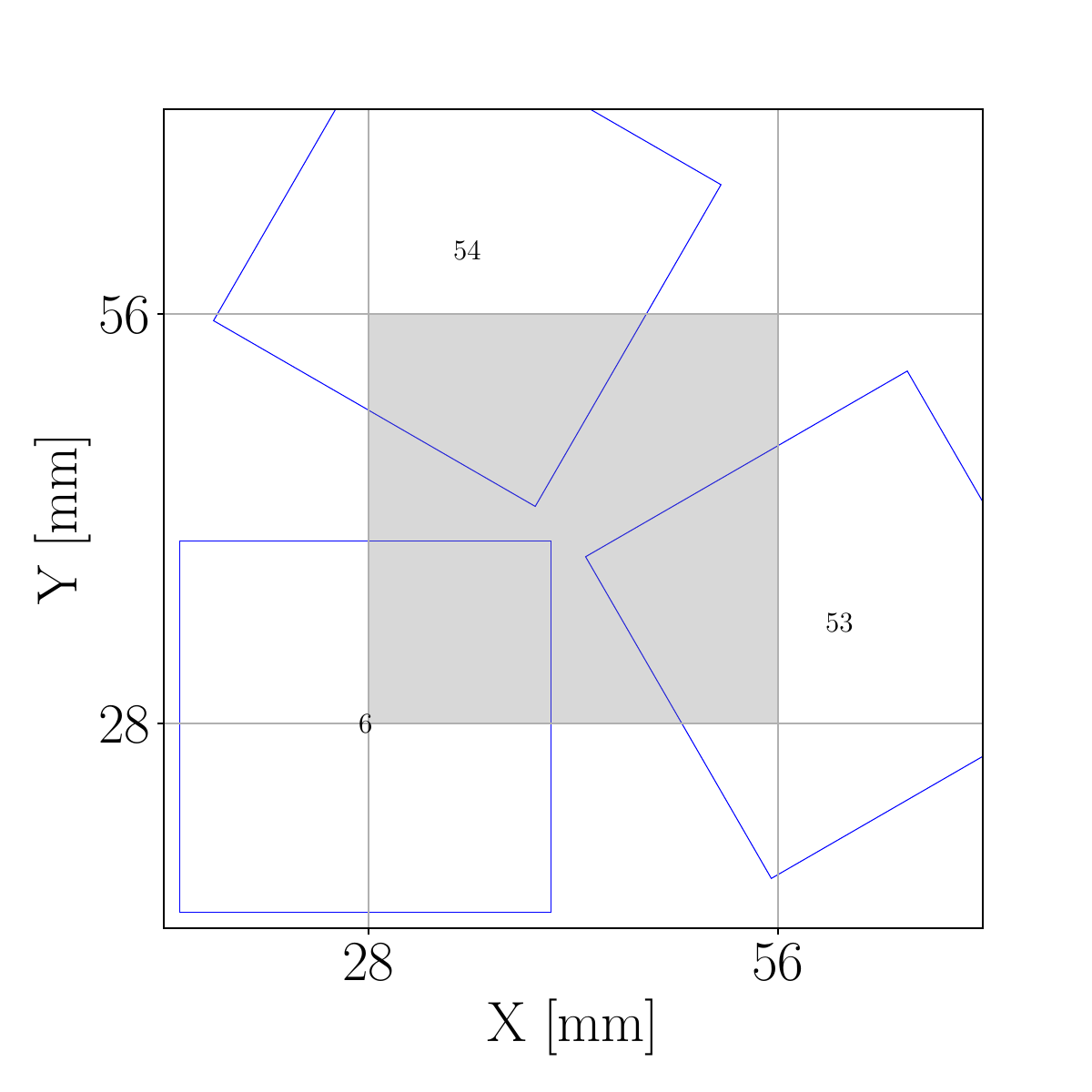}
    \caption{Illustration of the transformation from the 64-PMT array to a regular pixel grid. Each pixel value $V$ is computed as the sum of contributions from overlapping PMTs, weighted by the fractional overlap area $S_i$ between the pixel and the $i$-th PMT, and scaled by the normalized PMT signal $\eta_i$. The highlighted pixel region in the figure shows contributions from PMTs 6, 53, and 54, demonstrating how the geometrical overlap and relative signal intensities are combined to form the final pixel representation used as neural network input.}

    \label{fig:pixel}
\end{figure}


% \subsection{Datasets for Machine Learning}
% \label{subsec24}

% The input data to the network is a \(10 \times 10\) matrix filled with the signal strengths or relevant features from the PMTs , based on their positions and shapes. 
% The value of each pixel is calculated as:

% \begin{equation}   
% Value = \sum_{i=1}^{n} S_i \cdot \eta_i
% \end{equation}

% where \(S_i\) is the overlap area between the \(i\)-th PMT and the pixel, and \(\eta_i\) is the signal fraction of the \(i\)-th PMT among all top PMTs.

% The label for each input is the true 2D XY coordinates of the event, representing its actual position in the detector. The network learns the relationship between the PMT signal pattern and the true event location, enabling it to predict the position for new events based on their PMT signals.\autoref{fig:data_trans} shows how a pattern from a 64 PMT array is transformed into a \(10 \times 10\) matrix. 

\subsection{Training samples}
\label{subsec24}

To apply deep learning techniques for position reconstruction in the RELICS detector, the raw signals from the PMT array must be converted into a structured form suitable for input into convolutional neural networks. Due to the irregular spatial layout of the PMTs, we adopt a pixelation-based approach to encode the spatial distribution of light into a fixed-size image. This representation facilitates the learning of local spatial features essential for accurate regression of event positions.

The PMT array plane is divided into a \(10 \times 10\) grid. Each pixel value corresponds to the estimated fraction of light within its coverage, assuming a uniform distribution of the light signal across the PMT window. This can be expressed mathematically as:

\begin{equation}
\text{V} = \sum_{i=1} S_i \cdot \eta_i
\end{equation}

Here, V stands for one pixel value, \(S_i\) denotes the fractional overlap area between the \(i\)-th PMT and the pixel, representing the geometrical influence, while \(\eta_i\) is the normalized signal intensity of the \(i\)-th PMT, defined as the ratio of its measured photon count to the total photon count across all top PMTs for the event. This formulation ensures that both spatial and intensity information are preserved in the final representation. \autoref{fig:pixel} shows a paticular example of a pixel, and how its value is calculated. \autoref{fig:data_trans} illustrates an example of this transformation, showing how the photon pattern from a 64-PMT array is converted into a regular grid suitable for neural network input.

The ground truth label associated with each matrix is the true two-dimensional \((x, y)\) coordinate of the interaction vertex, as determined from the Monte Carlo simulation. The learning task is thus framed as a regression problem: given a signal distribution matrix, the network predicts the corresponding event position within the detector volume.

For the RELICS prototype detector, which consists of only 7 top PMTs, a smaller \(4 \times 4\) matrix is constructed using the same mapping procedure, adjusted to match the physical area covered by the PMTs. 

The training dataset comprises 80,000 simulated events, with an additional 20,000 events used for testing and 100,000 events for final validation. Each event corresponds to an average photon count of approximately 800 detected on the top array, representing typical scintillation signal intensities in the detector.

\section{Results and Discussion}
\subsection{Position reconstruction for RELICS}
In order to assess the feasibility of utilizing CycleGAN for DA in the RELICS experiment, we conduct a preliminary test using two distinct sets of simulation data. These datasets are generated with varying simulation parameters to represent different domain characteristics.

The primary objective of this test is to validate the ability of CycleGAN to learn and map the distributions of the two sets of data, thus enabling the transfer of knowledge from one domain to another. By leveraging CycleGAN's capability to learn complex mappings between paired datasets without the need for explicit labels, we aim to investigate its potential for bridging the gap between the data and simulation domains in the context of the RELICS detector.
Two sets of simulated data are generated to evaluate the performance of CycleGAN in the RELICS experiment:

\begin{table}[htbp]
\centering
\begin{tabular}{ccc}
\toprule
\textbf{Dataset} & \textbf{Signal Generation Method} & \textbf{Reflectivity of Teflon} \\
\midrule
$\text{MC}_0$ & Garfield++ & 0.99 \\
$\text{MC}_1$ & Point like & 0.70 \\
\bottomrule
\end{tabular}
\caption{Datasets used for training.}
\label{tab:datasets}
\end{table}

The primary difference between the two datasets lies in the method used for generating the scintillation signal and their respective reflectivity values. $\text{MC}_0$ is generated using input data provided by Garfield++ which stands for  more realistic scintillation signal simulation; $\text{MC}_1$ is generated using simple isotropic point-like light source to simulate scintillation signals. Also the reflectivity of the two datasets are set to different values (\autoref{tab:datasets}). $\text{MC}_0$ is treated as a hypothetical data domain, while $\text{MC}_1$ with a different distribution of scintillation light sources and reflectivity---serves as a simulation domain that intentionally differs from the real experimental environment.

% \begin{figure}[!htbp]
%     \centering
%     \begin{subfigure}[n]{\linewidth}
%         \centering
%         \includegraphics[width=\linewidth]{MC_1 (1).pdf}
%         \caption{A PMT pattern from $\text{MC}_1$.}
%         \label{fig:patternMC1}
%     \end{subfigure}
%     \vspace{0.1cm}
%     \begin{subfigure}[n]{\linewidth}
%         \centering
%         \includegraphics[width=\linewidth]{G_MC_1 (1).pdf}
%         \caption{PMT pattern from \autoref{fig:patternMC1} $\text{MC}_1$ after Domain Adaptation.}
%         \label{fig:patternGMC1}
%     \end{subfigure}
%     \vspace{0.1cm}
%     \begin{subfigure}[n]{\linewidth}
%         \centering
%         \includegraphics[width=\linewidth]{MC_0 (1).pdf}
%         \caption{A PMT pattern from $\text{MC}_0$.}
%         \label{fig:patternMC0}
%     \end{subfigure}
%     \caption{PMT pattern from different datasets. Red cross in the figure represent the real position of the event.The color of each PMT indicates the fraction of the total detected PE signal collected by that PMT within the top array.\autoref{fig:patternMC1} from the simulation domain is transformed into \autoref{fig:patternGMC1} using the generator from CycleGAN; \autoref{fig:patternMC0} is a sample from data domain. The generator done the transformation while keeping the feature of the original pattern. }
%     \label{fig:pattern}
% \end{figure}
\begin{figure}[!htbp]
    \centering
    \begin{subfigure}[t]{\linewidth}
        \centering
        \includegraphics[width=\linewidth]{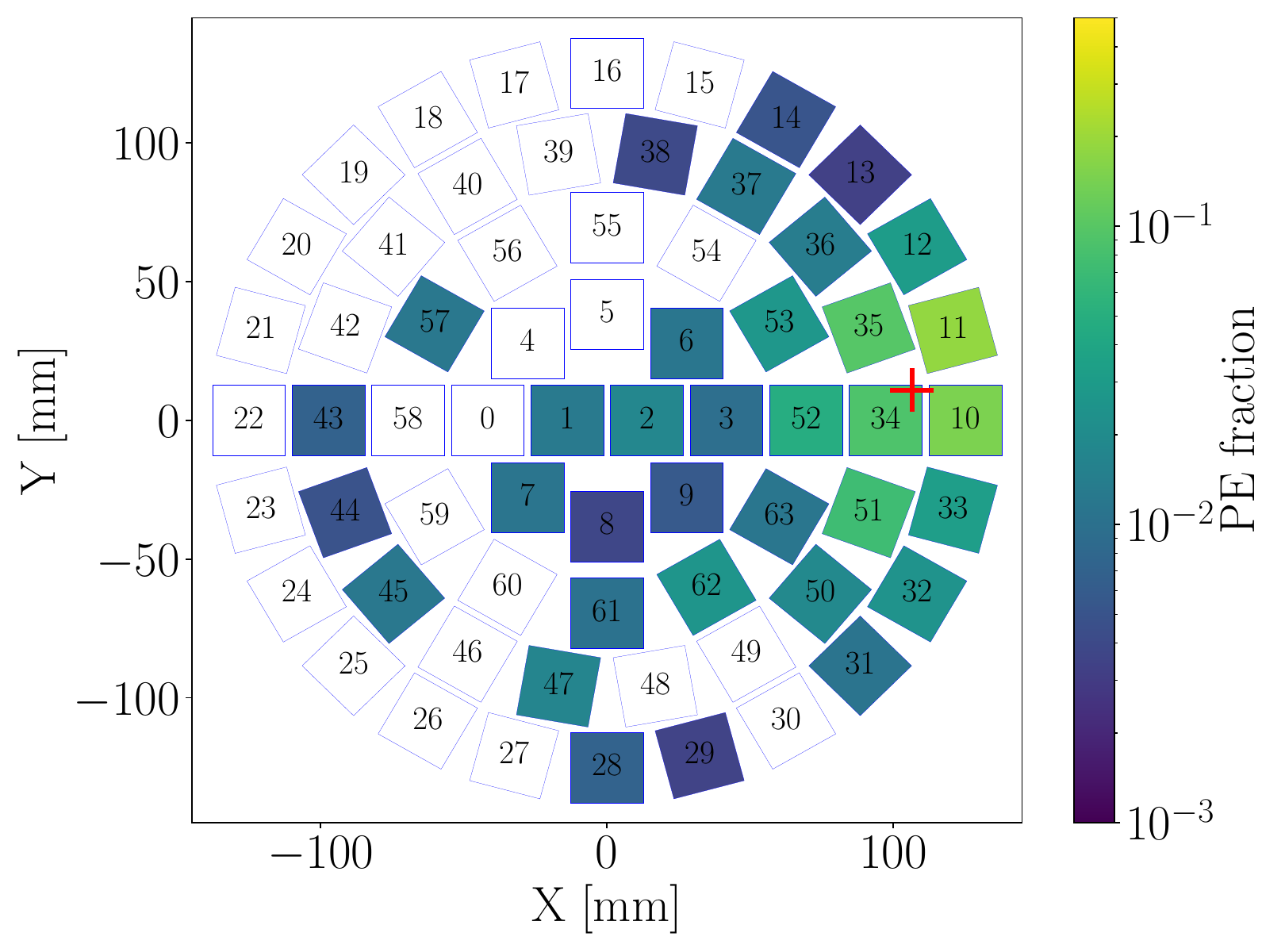}
        \caption{} % 留空
        \label{fig:patternMC1}
    \end{subfigure}
    \vspace{0.1cm}
    \begin{subfigure}[t]{\linewidth}
        \centering
        \includegraphics[width=\linewidth]{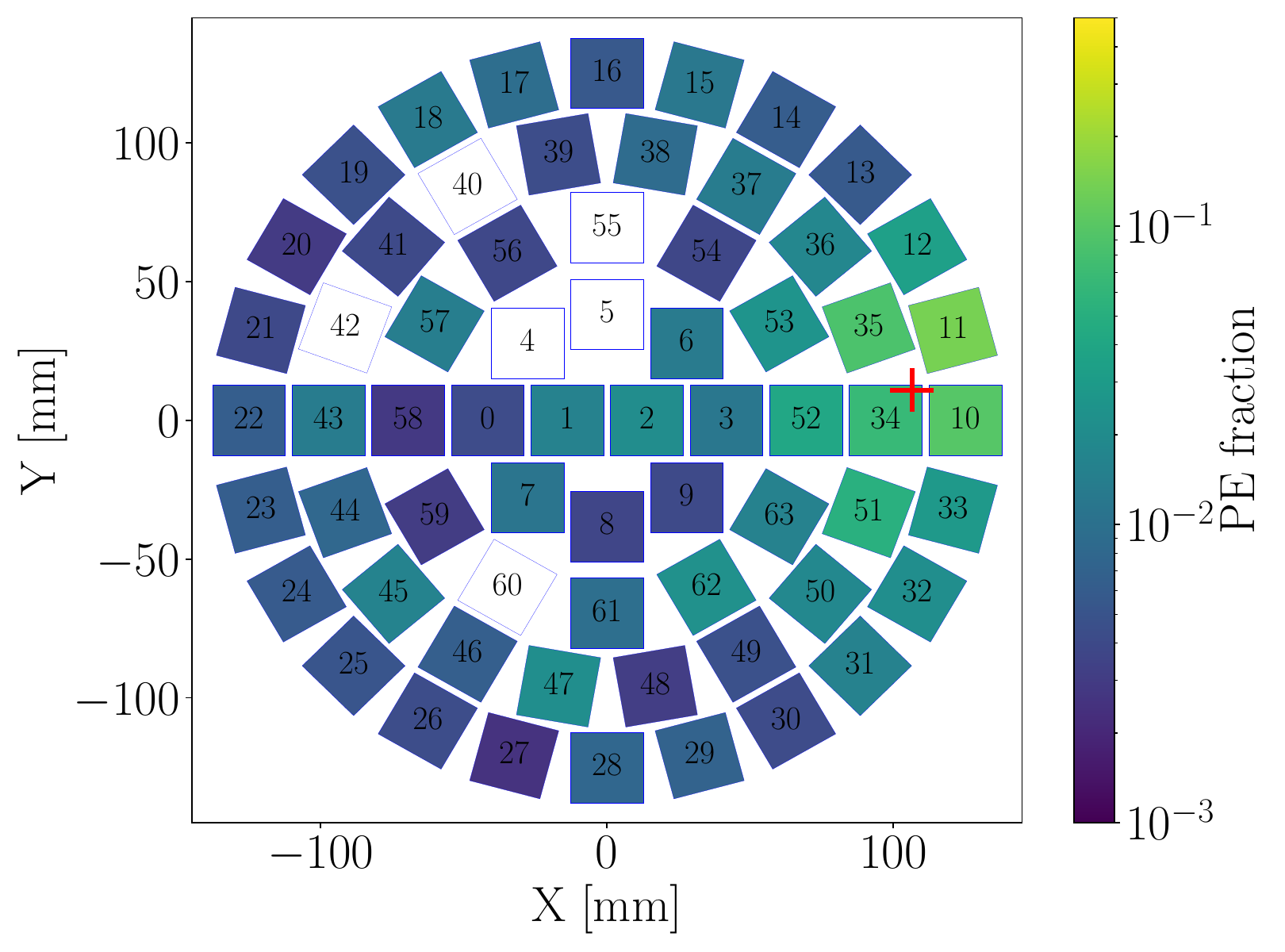}
        \caption{} % 留空
        \label{fig:patternGMC1}
    \end{subfigure}
    \vspace{0.1cm}
    \begin{subfigure}[t]{\linewidth}
        \centering
        \includegraphics[width=\linewidth]{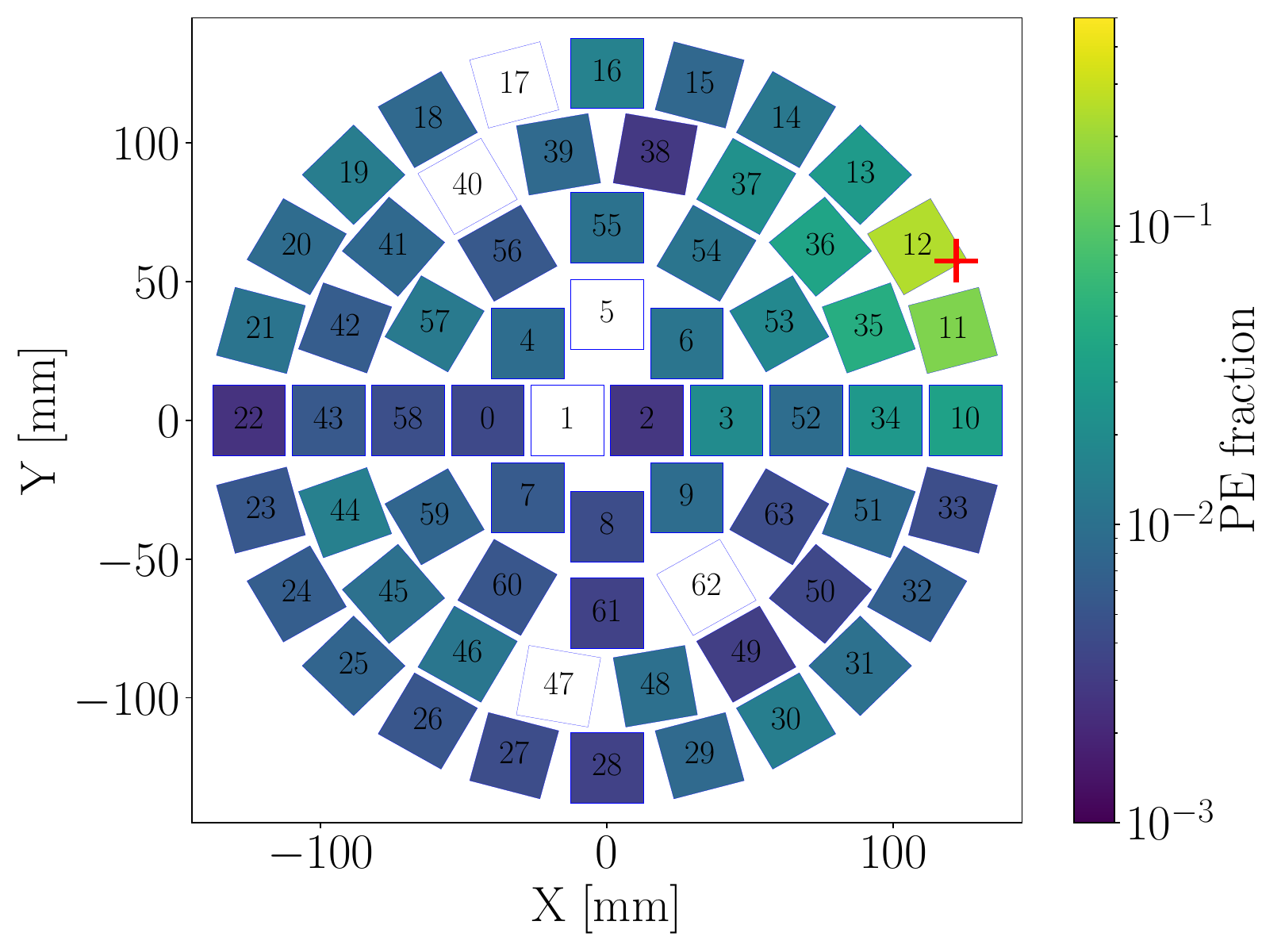}
        \caption{} % 留空
        \label{fig:patternMC0}
    \end{subfigure}
    \caption{
(a) PMT pattern from $\text{MC}_1$;
(b) PMT pattern from $\text{MC}_1$ after DA using the CycleGAN generator;  
(c) PMT pattern from $\text{MC}_0$ data domain.  
In all subfigures, the red cross indicates the true event XY position. The color of each PMT represents the fraction of total detected PE signal collected by that PMT in the top array. The generator performs the transformation while keeping the main features of the original pattern.}

    \label{fig:pattern}
\end{figure}

\begin{figure}[!htb]
    \centering
    \includegraphics[width=0.9\linewidth]{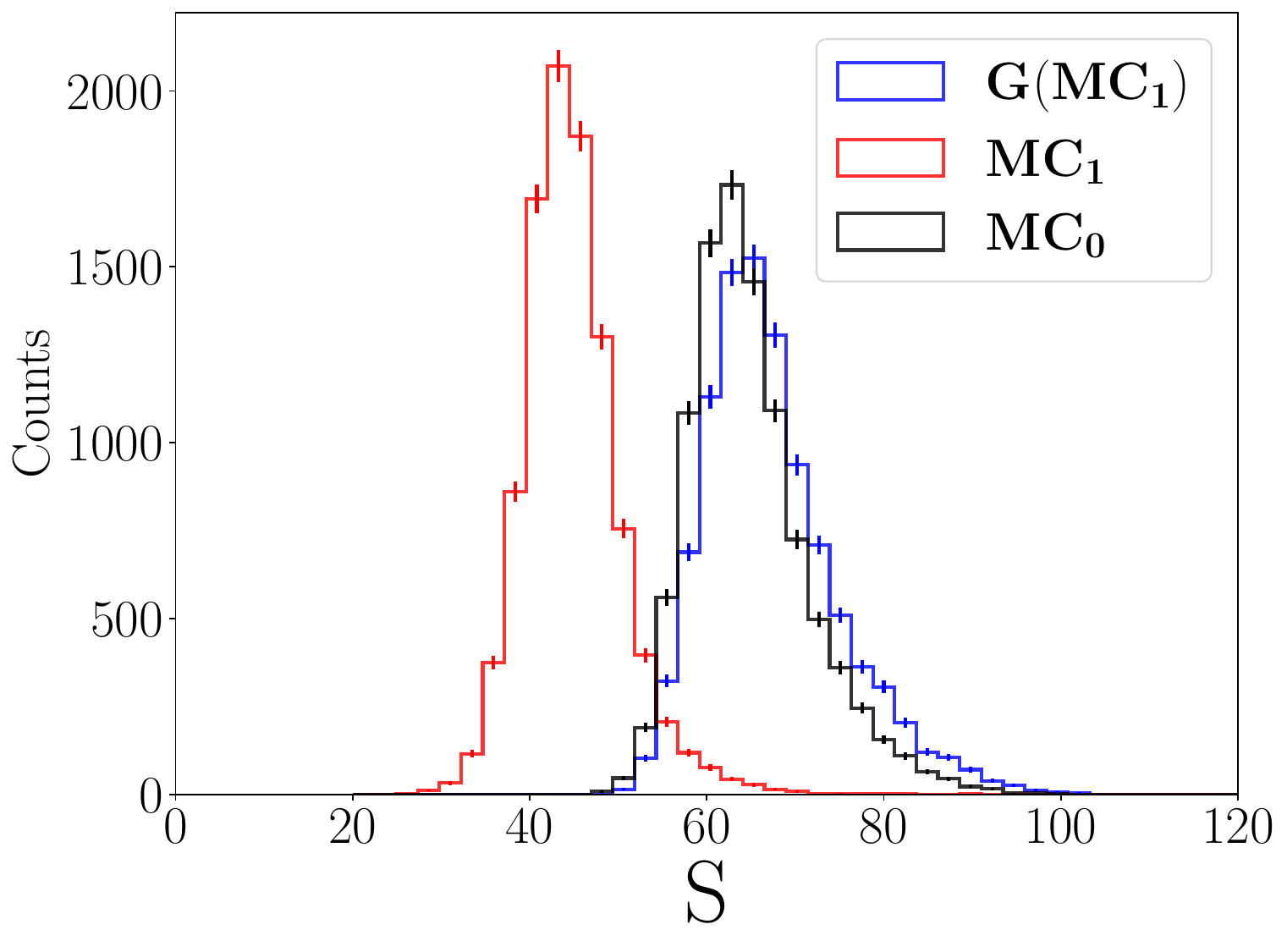}
    \caption{
Histogram of the spread of the top PMT patterns. A larger value indicates that the light detected by the top PMT array is more concentrated near the actual physical event position in the XY plane.
}

    \label{fig:spread}
\end{figure}

The results of the preliminary test demonstrate the effectiveness of CycleGAN for DA in the RELICS experiment. As shown in \autoref{fig:pattern}, the PMT patterns are more compatible between $\text{MC}_0$ and $\text{MC}_1$ after the generator G applied on $\text{MC}_1$. This transformation indicates that CycleGAN successfully adapts the simulation domain's data distribution to align more closely with that of the data domain.

% Further, as illustrated in \autoref{fig:spread}, the statistical measure of the spread of top PMT patterns reveals that, after domain adaptation, the distribution of the simulation domain becomes more similar to that of the data domain. This is confirmed by the improvement in the alignment of the statistical spread, highlighting the potential of CycleGAN to improve domain adaptation in the context of the RELICS experiment.

To evaluate the effect of DA on the spatial distribution of simulated data, we define a quantitative measure that captures the concentration of light detected by each PMT in the top array relative to the true XY position of the event. Specifically, for each top PMT, we compare its center position with the true event position, reflecting the correlation between the number of photons received by that PMT and the actual location of the charge deposition, since the events produce locally concentrated light, which is calculated by:

\begin{equation}
S={{\frac{\sum_{i=0}^{63} PE_i \cdot |\vec{r}_{i}-\vec{r}_{event}|}{\Sigma_{i=0}^{63} PE_i}}}
\end{equation}
Here, $i$ denotes the $i$-th top PMT ($i = 0,1,2,\dots,63$), $\vec{r}_i$ is the center position of the $i$-th PMT, and $\vec{r}_{event}$ is the true XY position of the charge signal.
As illustrated in \autoref{fig:spread}, after DA, the distribution of the simulation domain aligns more closely with that of the data domain. This improvement in alignment demonstrates that the CycleGAN effectively adjusts the simulated patterns to better match the experimental data, providing a clear and intuitive way to visualize the impact of DA on the simulated PMT responses.

As simulated data have known positions, we can directly calculate the difference between the reconstructed position \\$\vec{R}(x_{rec},y_{rec})$ and its true position $\vec{R}_0(x_{true},y_{true})$. 
The resolution is calculated with:
\begin{equation}
    |\vec{R}-\vec{R}_0| = \sqrt{(x_{rec}-x_{true})^2+(y_{rec}-y_{true})^2}
    \label{eq:totaldeviation}
\end{equation}

After performing DA using CycleGAN, we obtain a new dataset denoted as $G(\text{MC}_1)$. To evaluate the effect of DA, three ResNet models(All share the same architecture 
 in \autoref{fig:DRNarchitect}) are trained separately using \(\text{MC}_0\), \ \(\text{MC}_1\), and \ \(G(\text{MC}_1)\). All three models are then evaluated on a test set that follows the same distribution as \(\text{MC}_0\), and their resolution are calculated for comparison.

\begin{table}[!htb]
\centering
\renewcommand{\arraystretch}{1.3} % 增大行距
\begin{tabular}{lcc}
\toprule
\textbf{Training Set} & \textbf{resolution [mm]} & \textbf{1$\sigma$ Interval [mm]} \\
\midrule
$\text{MC}_0$ & $2.626$ & $[1.282,\, 4.688]$ \\
$\text{MC}_1$  & $3.913$ & $[1.797,\, 7.630]$ \\
$G(\text{MC}_1)$  & $2.857$ & $[1.364,\, 5.075]$ \\
\bottomrule
\end{tabular}
\caption{The resolution of the three models (100~$< R <$~140~mm, approximately 800 photons detected per event). The validation set is generated in the same way as $\text{MC}_0$.
}
\label{tab:resolution}
\end{table}

\autoref{tab:resolution} shows the resolution of three models. The resolution of the model trained on $\text{MC}_1$ data after DA ($G(\text{MC}_1)$) is $2.857$ mm, which demonstrates an 27 \% improvement compared to the resolution obtained without applying the DA model, $3.913$ mm.
It is important to note that for the central region (where $R<100$ mm), the reconstruction performances of all three models are quite similar. This is due to the fact that the patterns in this region are less influenced by the reflectivity factors. However, in the outer regions (where $R>100$ mm), the reconstruction performance varies more significantly, and DA has a much more pronounced effect in improving the accuracy. This is because the reflectivity factor plays a larger role in determining the pattern near the edges, and DA helps to correct for these discrepancies.

We observe radial reconstruction deviations near the edges of the TPC; however, since these occur outside our fiducial volume, they are discussed only in \autoref{appendix}.

% \begin{figure*}[!ht]
%     \centering

%     \begin{subfigure}[t]{0.48\linewidth}  % 左图宽度约占总宽的48%
%         \centering
%         \includegraphics[width=\linewidth]{resolution.pdf}
%         \caption{resolution $|\vec{R} - \vec{R}_0|$ vs. $\vec{R}_0$}
%         \label{fig:total_deviation}
%     \end{subfigure}
%     \hfill  % 两图之间的水平间距，可以用 \hspace{<长度>} 替换
%     \begin{subfigure}[t]{0.48\linewidth}  % 右图宽度约占总宽的48%
%         \centering
%         \includegraphics[width=\linewidth]{deviation.pdf}
%         \caption{Radial deviation $R - R_0$ vs. $R_0$}
%         \label{fig:R_R0}
%     \end{subfigure}

%     \caption{The deviation between reconstructed and true positions as a function of the true radius. The blue dots represent the median deviation. $\vec{R}_0$ is the true position, $\vec{R}$ is the reconstructed position; $R_0$, $R$ represent the true and reconstructed radius, respectively.}
%     \label{fig:deviation}
% \end{figure*}

\subsection{Test on RELICS prototype}
% \begin{figure*}[!ht]
%     \centering
%     \begin{subfigure}[t]{0.48\linewidth}
%         \centering
%         \includegraphics[width=\linewidth]{protoOrigin.pdf}
%         \caption{Using model trained by Simulated Data}
%         \label{fig:proto_origin}
%     \end{subfigure}
%     \hfill
%     \begin{subfigure}[t]{0.48\linewidth}
%         \centering
%         \includegraphics[width=\linewidth]{proto_after_da.pdf}
%         \caption{Using model trained by Simulated Data after Domain Adaptation}
%         \label{fig:proto_da}
%     \end{subfigure}

%     \caption{Difference in the distribution of reconstructed events using two models. The red crosses represent the geometric centers of the top PMTs. The color indicates the number of events in each 2D bin.}
%     \label{fig:two_images}
% \end{figure*}
For the RELICS prototype operated at Tsinghua University, we have already collected a sufficient number of real physics events with corresponding PMT signals. $^{37}\mathrm{Ar}$ is a commonly used calibration source in liquid xenon detectors. After being injected into the gas circulation system, the $^{37}\mathrm{Ar}$ atoms are carried into the main detector volume as the xenon gas is liquefied through the cryogenic system. Therefore, the distribution of this calibration source in the liquid xenon is generally considered to be uniform. Also, the energy of this calibration source is very close to that of our region of interest (ROI).
% \begin{figure*}[!h]
%     \centering
%     \begin{subfigure}[n]{0.48\linewidth}
%         \centering
%         \includegraphics[width=1\linewidth]{protoOrigin.pdf}
%         %\caption{Using model trained by Simulated Data}
%         \caption{}
%         \label{fig:proto_origin}
%     \end{subfigure}
%     % \vspace{0.1cm} % 添加垂直间距
%     \begin{subfigure}[n]{0.48\linewidth}
%         \centering
%         \includegraphics[width=1\linewidth]{proto_after_da.pdf}
%         %\caption{Using model trained by Simulated Data after DA}
%         \caption{}
%         \label{fig:proto_da}
%     \end{subfigure}
%     \caption{Distribution of reconstructed events using two models, (a) uses model trained by simulation domain, (b) uses model trained by simulation domain after DA. The red crosses represent the geometric centers of the top PMTs. The color indicates the number of events in each 2D bin. }
%     \label{fig:two_images}
% \end{figure*}

\begin{figure}[!h]
    \centering
    \begin{subfigure}[n]{0.9\linewidth}
        \centering
        \includegraphics[width=1\linewidth]{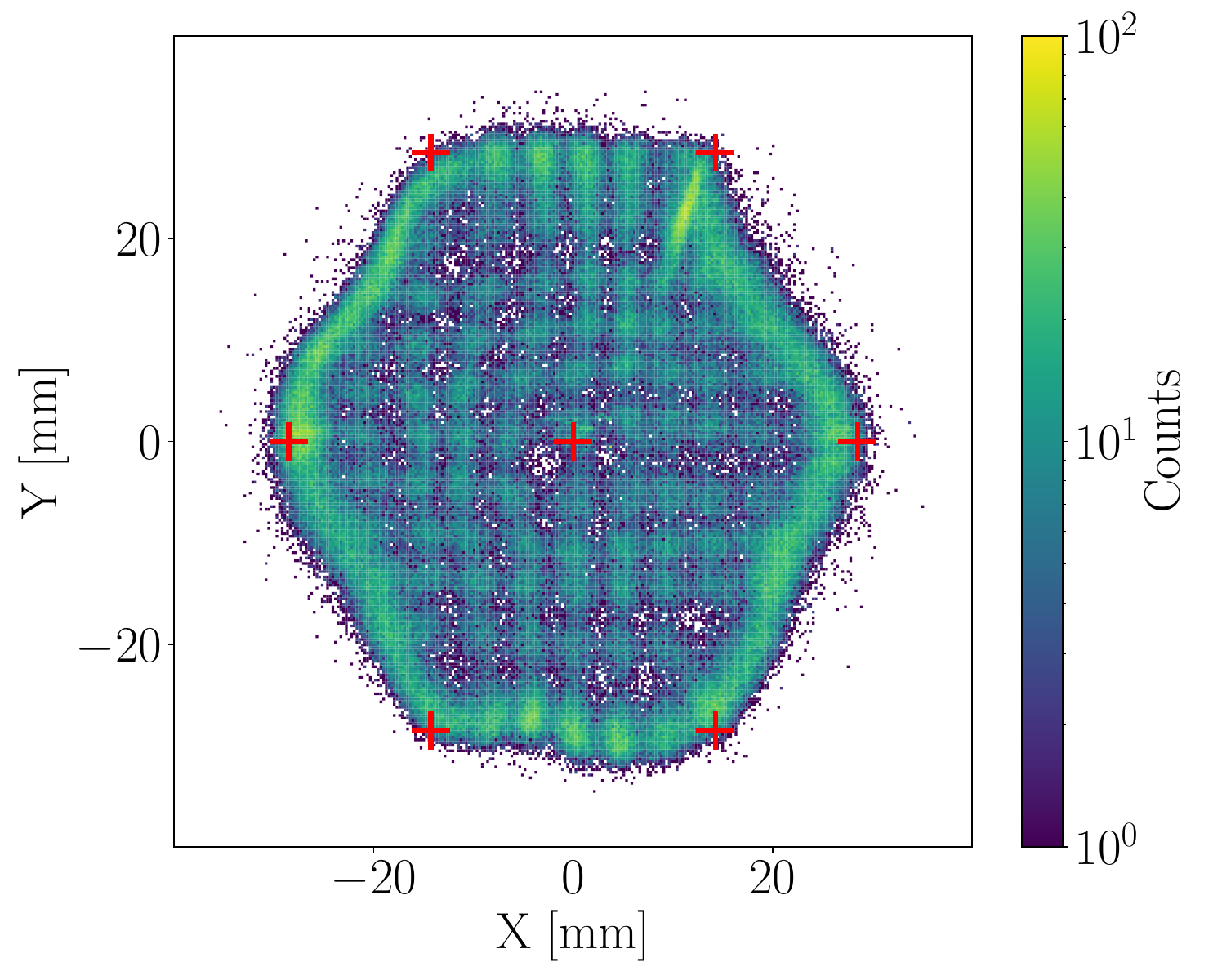}
        %\caption{Using model trained by Simulated Data}
        \caption{}
        \label{fig:proto_origin}
    \end{subfigure}
    % \vspace{0.1cm} % 添加垂直间距
    \begin{subfigure}[n]{0.9\linewidth}
        \centering
        \includegraphics[width=1\linewidth]{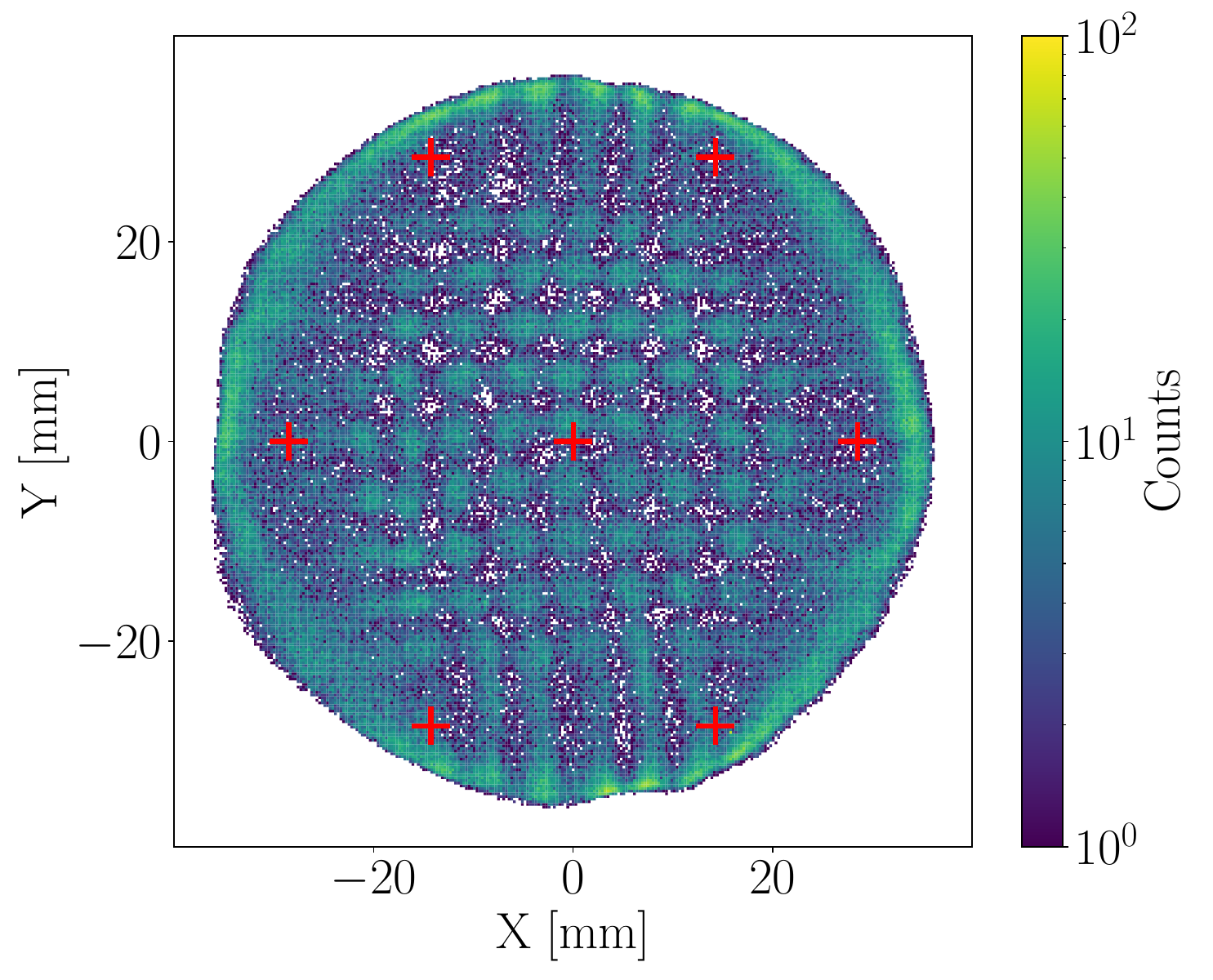}
        %\caption{Using model trained by Simulated Data after DA}
        \caption{}
        \label{fig:proto_da}
    \end{subfigure}
    \caption{Distribution of reconstructed events using two models, (a) uses model trained by simulation domain, (b) uses model trained by simulation domain after DA. The red crosses represent the geometric centers of the top PMTs. The color indicates the number of events in each 2D bin. }
    \label{fig:two_images}
\end{figure}

\begin{figure}[!h]
    \centering
    \includegraphics[width=0.9\linewidth]{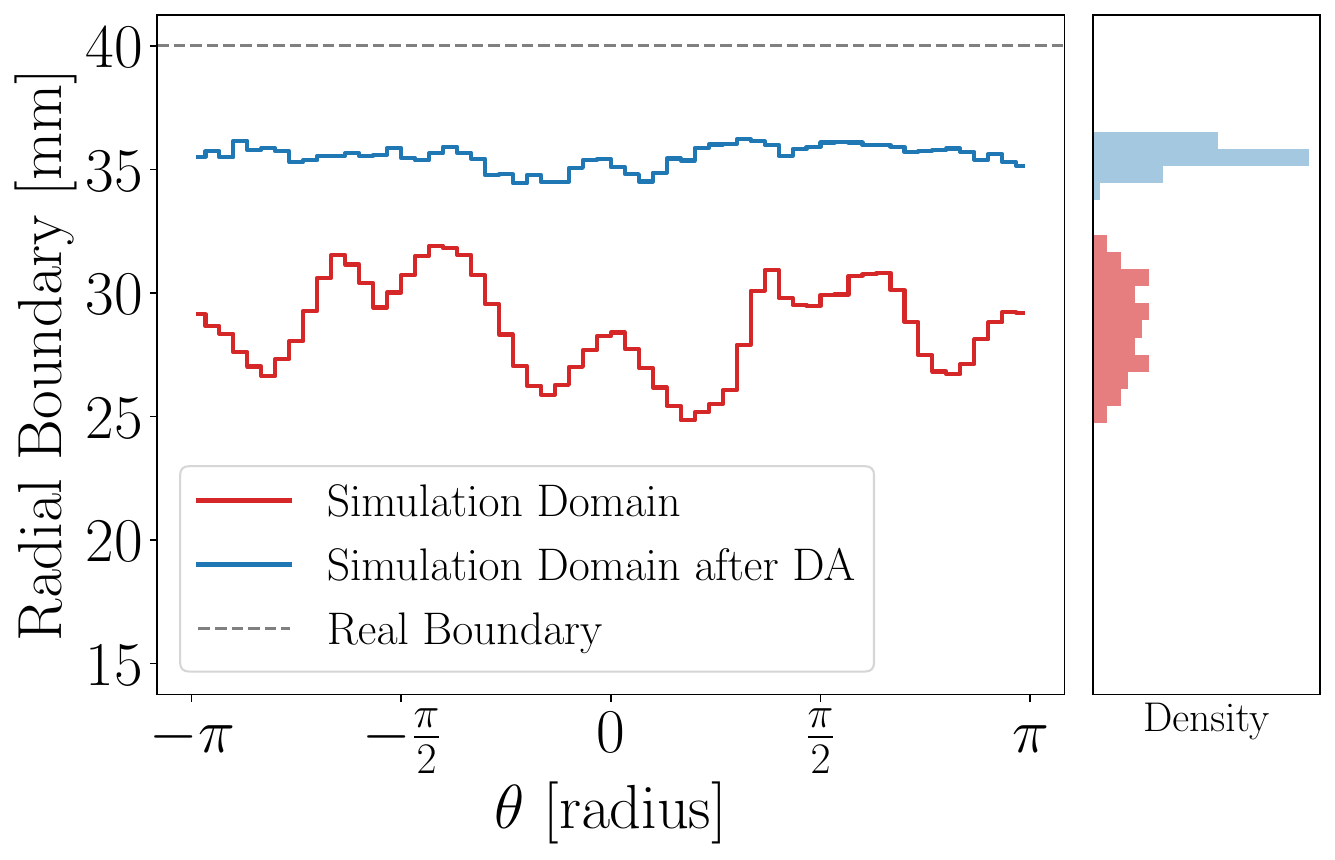}
    \caption{
Radial boundary \( R \) versus polar angle \( \theta \). The red and blue step lines represent the reconstruction of ResNet trained with Simulated Data and Simulated Data after DA , respectively. The dashed gray line shows the reference circular boundary at 40 mm. The histogram on the right illustrates the radial distribution of both models. 
}
    \label{fig:boudary}
\end{figure}

The two X-Y distributions shown in \autoref{fig:two_images} illustrate the difference in the reconstruction of $^{37}\mathrm{Ar}$ signals using two models: one trained with simulation domain (\autoref{fig:proto_origin}) and the other trained with simulation domain after DA, which should be closer to the data domain (\autoref{fig:proto_da}).100,000 events for each data domain and simulation domain are used to tune CycleGAN. The X-Y distributions seem to manifest as isolated block clusters, which is caused by the electric field focusing electrons toward the center of the electrode grid during their drift (Also shown in the simulation: \autoref{fig:Garfield++}). In our case the distance between cluster centers matches the the pitch of the electrode grid (5 mm). Additionally, the model trained solely on the simulation domain shows poor reconstruction performance near the edges, while the model trained after DA exhibits a distribution that more closely matches the cylindrical boundary ($R = 40$ mm). The accumulation near the boundary indicates the presence of bias close to the TPC edge, suggesting that the DA algorithm does not effectively mitigate this issue.

% \begin{figure}[!htb]
%     \centering
%     \includegraphics[width=\linewidth]{R2.pdf}
%     \caption{$R^2$ distribution of reconstructed events using two models.The dashed lines in the figure indicate the boundary of the flat region for each reconstruction model.}
%     \label{fig:R2distribution}
% \end{figure}

As shown in Figure~\ref{fig:boudary}, the radial boundary \( R \) at each polar angle \( \theta \) is defined as the radius within which 95\% of points lie in each angular sector. This approach captures the angular-dependent shape of the point clouds, allowing a comparison between models before and after domain adaptation. The comparison of these boundaries reveals a significant improvement after DA: the mean deviation relative to the ideal circular boundary of 40 mm decreases from 11.36 mm in the model before DA to 4.47 mm, a standard deviation of 0.46 mm (before DA: 1.876 mm), indicating a substantial enhancement in boundary reconstruction accuracy. Furthermore, the standard deviation of the error is reduced, reflecting increased consistency across angular sectors. Compared to the model before DA, the domain-adapted model reduces the mean absolute boundary deviation, resulting in a 60.6\% improvement in reconstruction accuracy.

% \newpage

\section{Conclusion}

\begin{figure}[htb]
    \centering
    \includegraphics[width=0.9\linewidth]{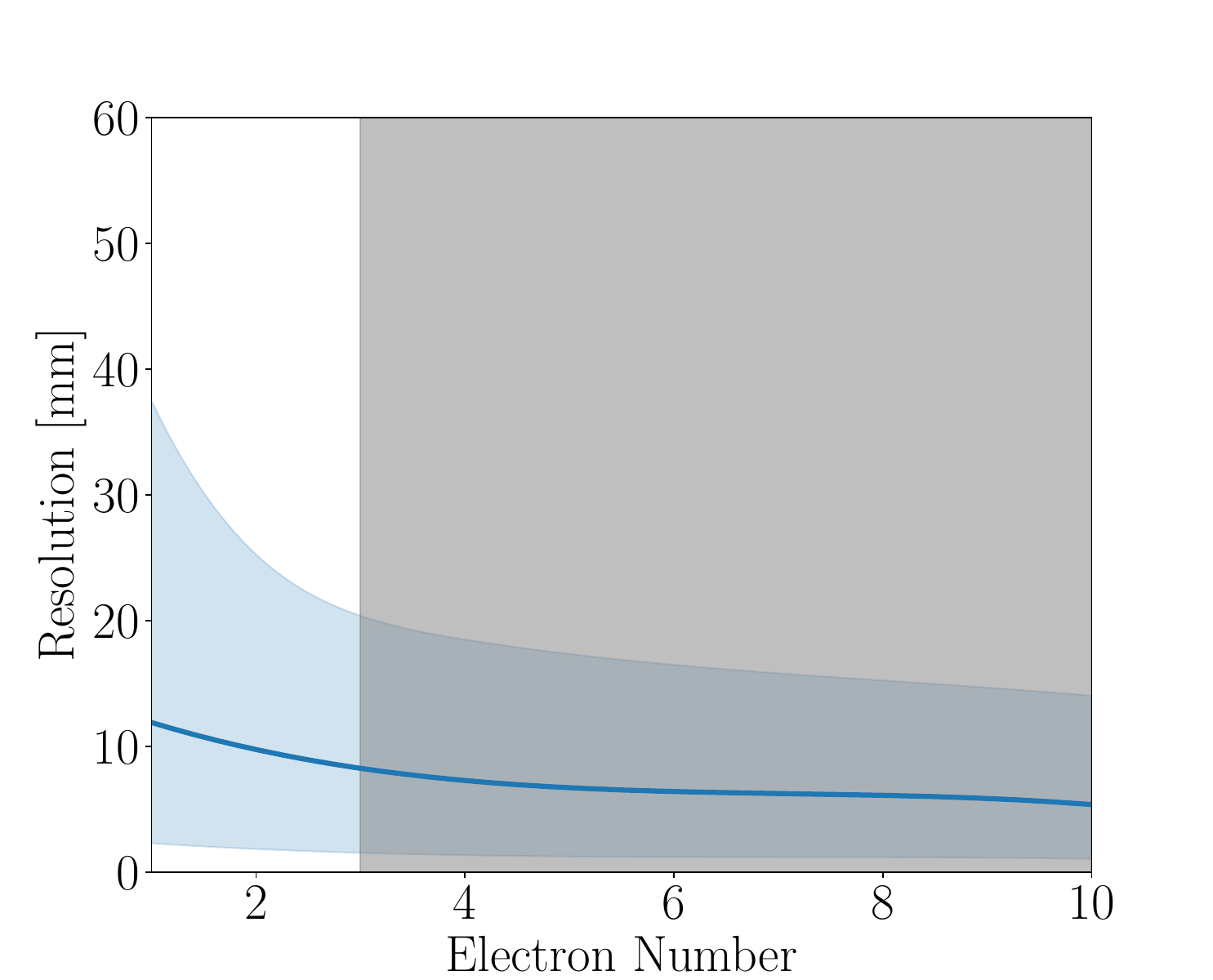}
    \caption{Reconstructed resolution VS Electron Number  liberated and extracted into gas xenon  (Assume $ QE = 30\% $ , Fluctuation the Gain of each PMT follows: $N(1,0.265^2)$. The shaded area represents the ROI for searching CE$\nu$NS signal.}
    \label{fig:resolutionROI}
\end{figure}
This study highlights the effectiveness of CycleGAN-based DA in enhancing position reconstruction in TPCs, especially in addressing domain shifts and model mismatches between simulated and real-world data.
In the prototype detector, the boundary reconstruction accuracy is improved by 60.6\% after applying the domain-adapted model, the detailed development of RELICS prototype will be published\citep{relicsprototype}. Applying this approach to RELICS experiment, which will be a 50-kg TPC, we achieved a spatial resolution of \(6.478 \ \mathrm{mm}\) (\([3.106, 11.117] \ \mathrm{mm}\) as the \(1\sigma\) region) within the region of interest (ROI) corresponding to 3–10 electron events. The resolution is shown as a function of the event's electron number in \autoref{fig:resolutionROI}.

CycleGAN-based DA not only offers an effective solution for enhancing position reconstruction in dual-phase TPCs, but also can be used in other pattern-based reconstruction algorithms.  The performance shown in our experiments indicates this is a worthwhile path to explore. 

\section*{Acknowledgements} We thank the RELICS collaboration for the fruitful discussions. This project is supported in part by grants from the Natural Science Foundation of China (No. 12090060, No. 12090061, No. 12275267, No. 12250011), by a grant from the Ministry of Science and Technology of China (No. 2021YFA1601600), and by the Frontier Scientific Research Program of Deep Space Exploration Laboratory Grants (No. 2022-QYKYJH-HXYF-013 and No. QY02ZHL001ZC3KX-2426).
\appendix

\section{Radial deviation}
\label{appendix}
We estimate the uncertainties of the reconstruction using the simulated data. Besides the resolution (\autoref{eq:totaldeviation}), we calculate the radial deviation with:
\begin{equation}
    |\vec{R}|-|\vec{R}_0| = \sqrt{x_{rec}^2+y_{rec}^2}-\sqrt{x_{true}^2+y_{true}^2}
\end{equation}

\begin{figure}[!htb]
    \centering
    \begin{subfigure}[n]{0.8\linewidth}
        \centering
        \includegraphics[width=1\linewidth]{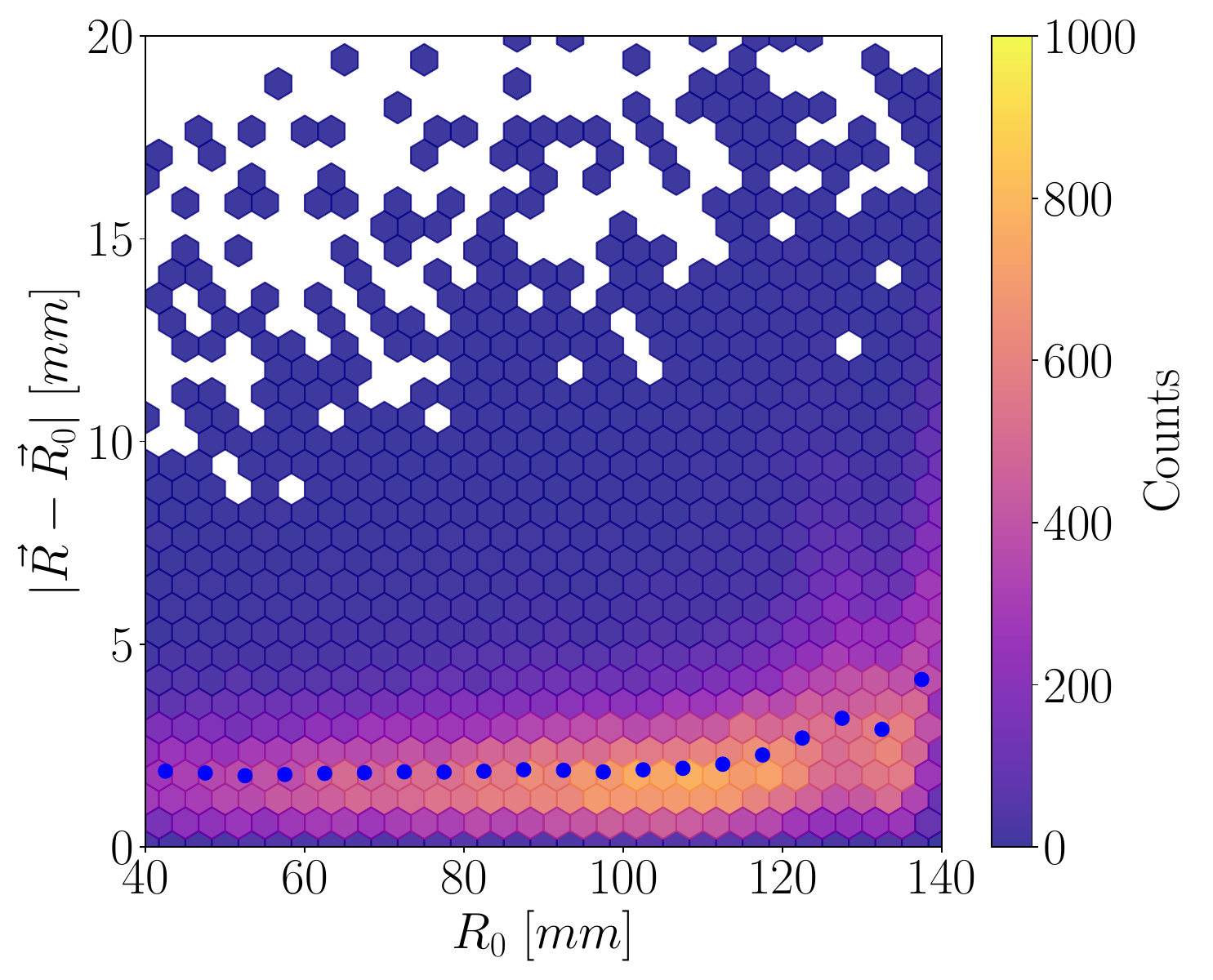}
        \caption{}
        \label{fig:resolution}
    \end{subfigure}
    \begin{subfigure}[n]{0.8\linewidth}
        \centering
        \includegraphics[width=1\linewidth]{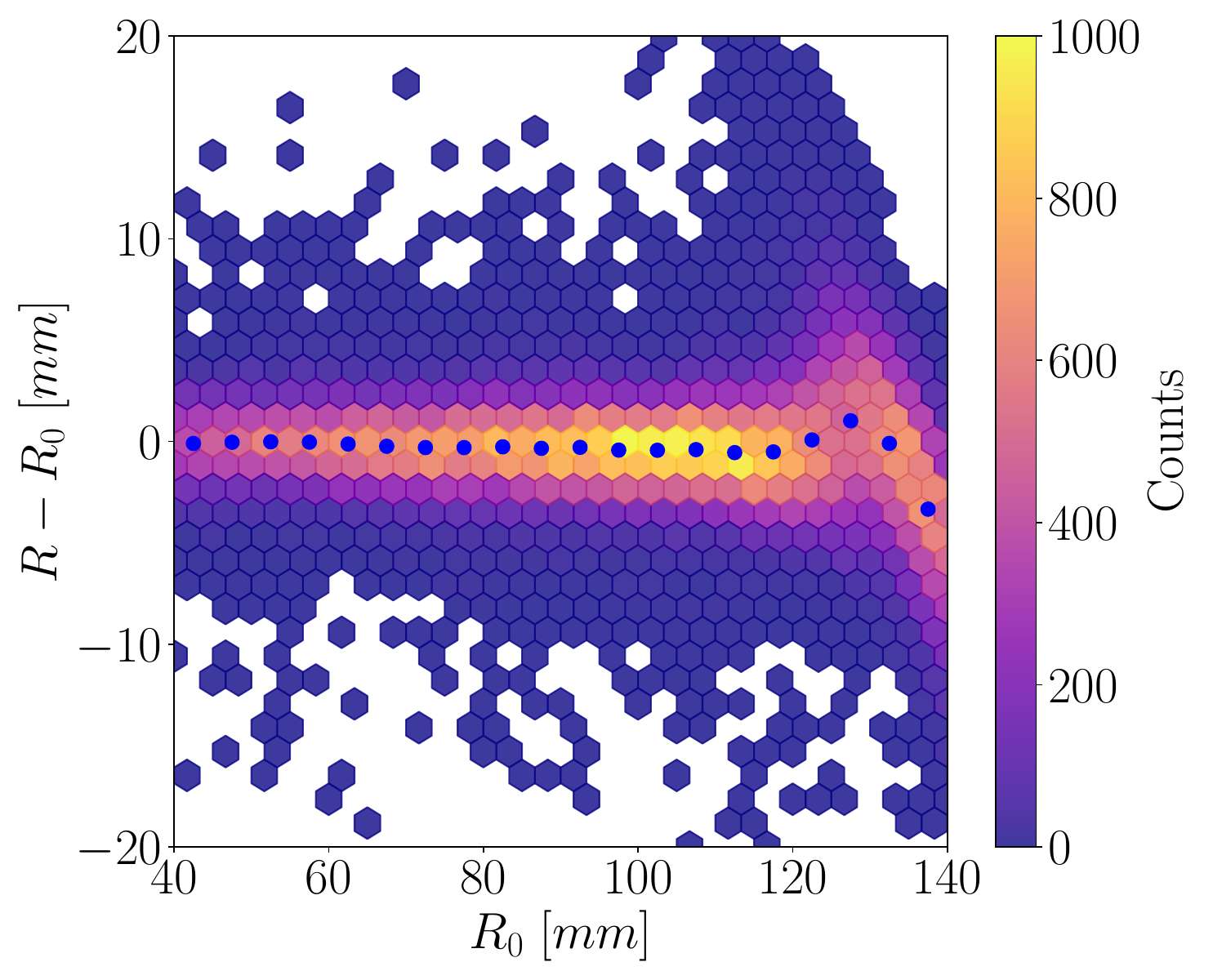}
        \caption{}
        \label{fig:R-R0}
    \end{subfigure}

    \caption{(a) Resolution $|\vec{R} - \vec{R}_0|$ vs. $\vec{R}_0$; (b) Radial deviation $R - R_0$ vs. $R_0$. The blue dots represents the median deviation. $\vec{R}_0$ is the true position, $\vec{R}$ is the reconstructed position; $R_0$, $R$ represents the ture and reconstructed radius.   }
    \label{fig:deviation}
\end{figure}

Those two kinds of deviation are plotted in \autoref{fig:deviation} as a function of the true radius $R_0$. We can observe that the reconstruction near the edge of the TPC performs worse.

We expect the resolution to increase as the edge approaches due to fewer PMTs that constitute the main feature of the pattern. What makes us concern is the deviation in the radial direction.We believe that there are two main reasons for explaining the radial deviation: (i) Reflections caused by Teflon. (ii) Intrinsic uncertainties of Machine Learning method.

\begin{figure}[!h]
    \centering
    \includegraphics[width=0.8\linewidth]{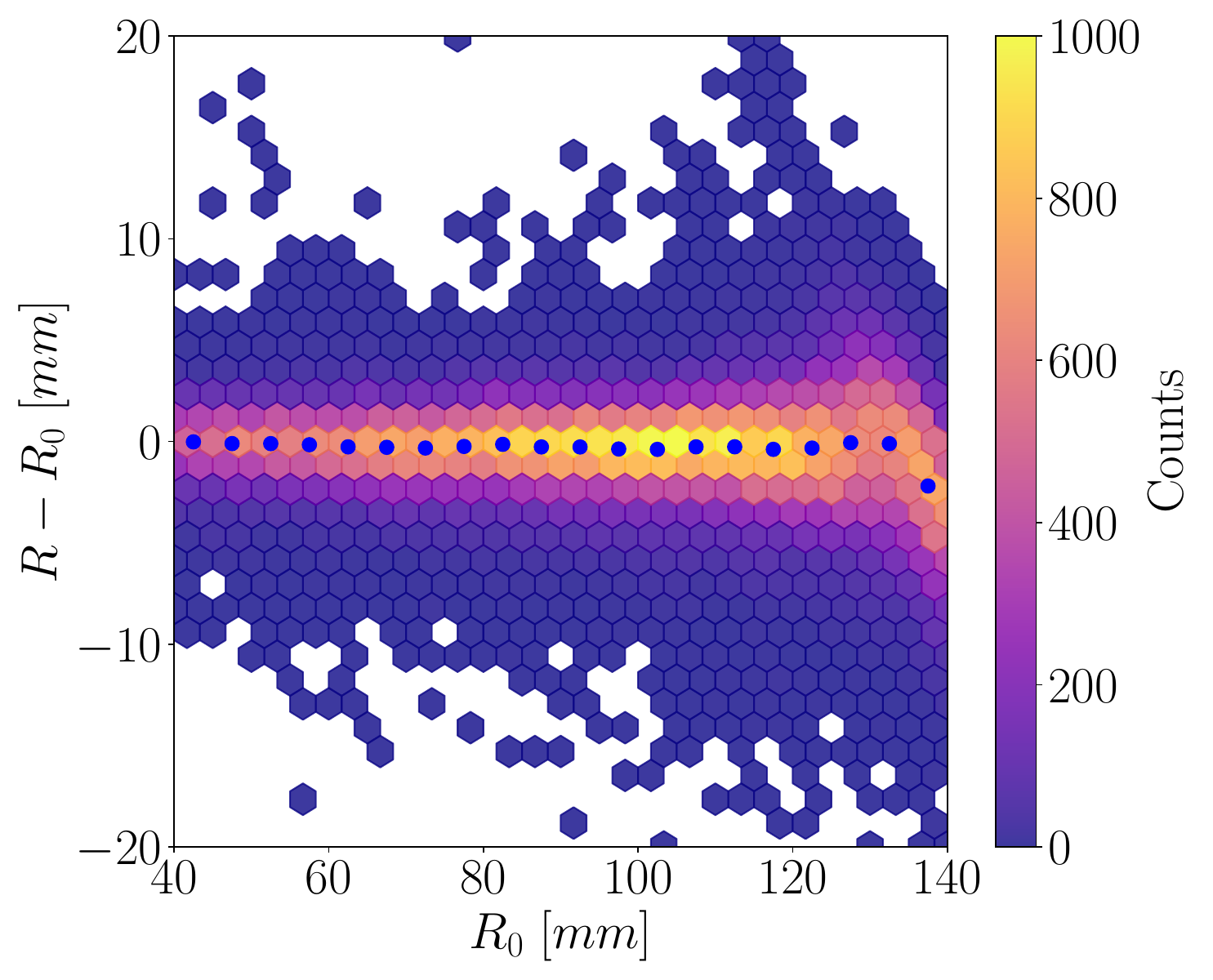}
    \caption{Radial deviation as a function of $R_0$, trained and validated with simulated data which forced the reflectivity of Teflon to 0. }
    \label{fig:deviation0}
\end{figure}

\begin{figure}[!h]
    \centering
    \includegraphics[width=0.8\linewidth]{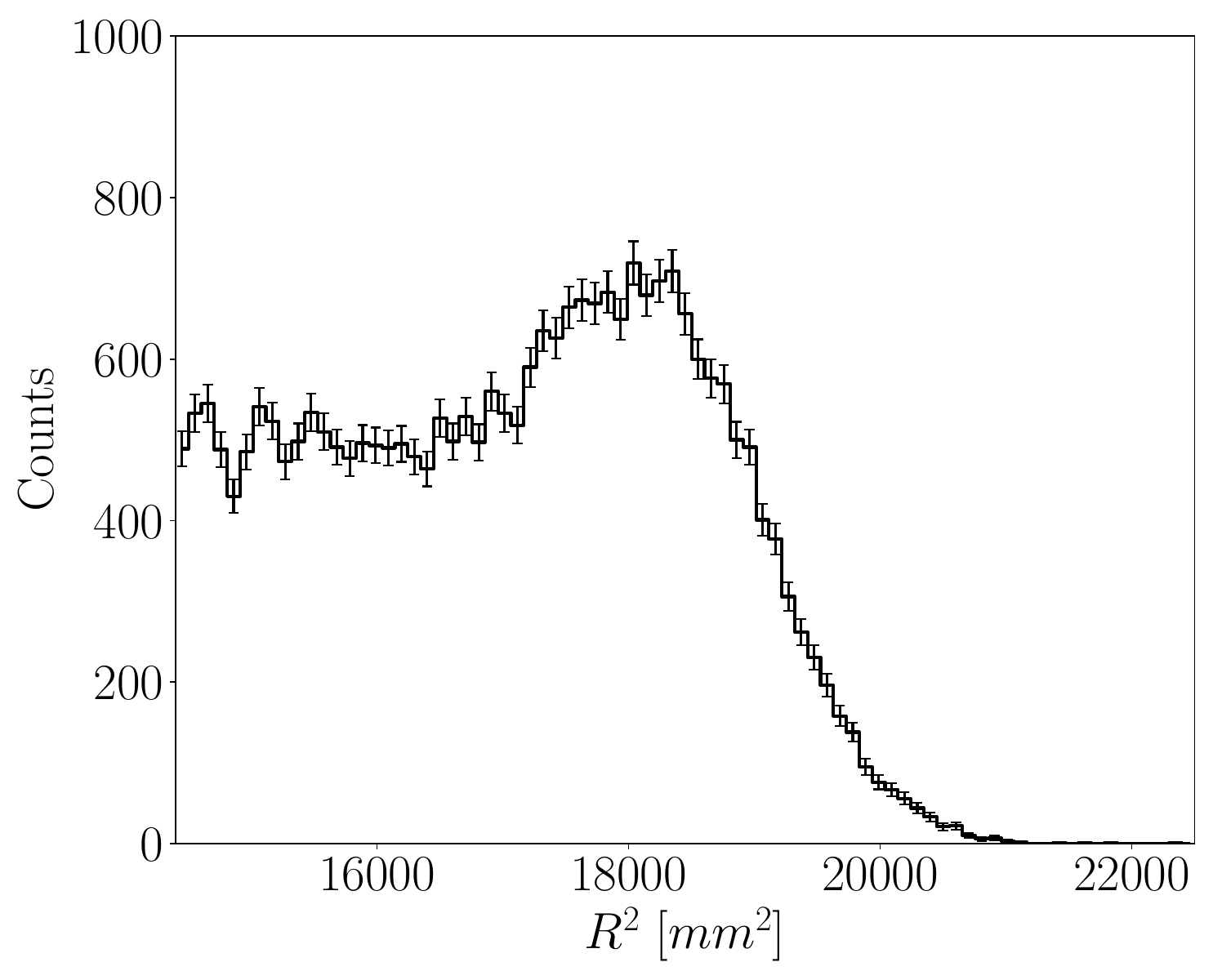}
    \caption{$R^2$ distribution of an uniform distribution validation set, R is the reconstructed radius. This can explain, to some extent, the accumulation near the boundary shown in  \ref{fig:proto_da}  }
    \label{fig:R2distribution}
\end{figure}
To isolate the component of the radial deviation attributed to Teflon reflections, we manually set the Teflon reflectivity to zero. As shown in \autoref{fig:deviation0}, the corresponding radial deviation is displayed. In comparison with \autoref{fig:deviation}, the noticeable “bump” around \( R_0 \in (120, 130)\,\mathrm{mm} \) disappears, indicating that this feature is caused by the contribution of Teflon reflections.

\autoref{fig:deviation0} still shows a minus deviation near the edge, which means there will be an inward bias. Those bias will cause an accumulation around $R \approx 135 \mathrm{mm}$ area, which can be observed in \autoref{fig:R2distribution}. This deviation is due to the fact that within this range, the change in the Pattern with respect to \( R \) is not significant, making it difficult for the neural network to learn this part of the data. This deviation can only be reduced by modifying the arrangement of the PMT array to increase its sensitivity to changes in \( R \) near the TPC edge, or by using a more powerful network. As our current Fiducial Volume $R_{FV}\in [0,120] \mathrm{mm} $, the mitigation of this deviation is not urgent.

\bibliographystyle{elsarticle-num}
\bibliography{./ref.bib}     
\end{document}